\documentclass[onecolumn,nobibnotes,nofootinbib,superscriptaddress]{revtex4}
\usepackage{amsmath,amssymb}
\usepackage[a4paper,bindingoffset=0.2in,left=0.8in,right=0.8in,top=1in,bottom=1in,footskip=.25in]{geometry}
\usepackage{graphicx,subfigure,epsfig}
\usepackage{bigints}
\usepackage{multirow}

\newcommand{\be}{\begin{equation}}
\newcommand{\ee}{\end{equation}}
\newcommand{\bea}{\begin{eqnarray}}
\newcommand{\eea}{\end{eqnarray}}
\newcommand{\benn}{\begin{displaymath}}
\newcommand{\eenn}{\end{displaymath}}
\newcommand{\beann}{\begin{eqnarray*}}
\newcommand{\eeann}{\end{eqnarray*}}
\newcommand{\abar}{\bar{\alpha}_s}

\begin{document}
\title{Exclusive photoproduction of vector meson at next-to-leading order from Color Glass Condensate}
\author{Yanbing Cai}
\email{yanbingcai@mail.gufe.edu.cn}
\affiliation{Guizhou Key Laboratory in Physics and Related Areas, Guizhou University of Finance and Economics, Guiyang 550025, China}
\author{Wenchang Xiang}
\email{wxiangphy@gmail.com}
\affiliation{Guizhou Key Laboratory in Physics and Related Areas, Guizhou University of Finance and Economics, Guiyang 550025, China}
\affiliation{Department of Physics, Guizhou University, Guiyang 550025, China}
\author{Mengliang Wang}
\email{mengliang.wang@mail.gufe.edu.cn}
\affiliation{Guizhou Key Laboratory in Physics and Related Areas, Guizhou University of Finance and Economics, Guiyang 550025, China}
\author{Daicui Zhou}
\email{dczhou@mail.ccnu.edu.cn}
\affiliation{Key Laboratory of Quark and Lepton Physics (MOE) and Institute of Particle Physics, Central China Normal University, Wuhan 430079, China}


\begin{abstract}
The exclusive photoproduction of vector mesons ($J/\psi$ and $\phi$) are investigated by taking into account the next-to-leading order corrections in the framework of Color Glass Condensate. We confront the next-to-leading order modified dipole amplitude with the HERA data finding good agreement. Our studies show that the $\chi^2/d.o.f$ from leading order, running coupling and collinearly improved next-to-leading order dipole amplitudes are 2.159, 1.097, and 0.932 for the elastic cross section, and 2.056, 1.449, and 1.357 for differential cross section. The outcomes indicate that the higher-order corrections have a significant contribution to the vector meson productions and the description of the experimental data is dramatically improved once the higher order corrections are included. We extend the next-to-leading order exclusive vector meson production model to LHC energies by using the same parameters obtained from HERA. We find that our model can also give a rather good description of the $J/\psi$ and $\phi$ data in proton-proton collision at 7 TeV and 13 TeV at LHCb experiments.
\end{abstract}

\maketitle


\section{Introduction}
Perturbative Quantum Chromodynamics (pQCD) predicts that the gluon density inside a hadron grows rapidly with increasing energy (or decreasing Bjorken-$x$) and saturates eventually at sufficiently high energies, forming a new state of high density gluonic matter called Color Glass Condensate (CGC)\cite{IV}. The rapidity evolution of the CGC matter is known to be described by Balitsky-JIMWLK\footnote{The JIMWLK is the abbreviation of Jalilian-Marian, Iancu, McLerran, Weigert, Leonidov, Konver} equation\cite{JKLW1,JKLW2,ILM,FILM} whose mean field version is called Balitsky-Kovchegov (BK) equation\cite{Balitsky,Kovchegov}. One of the hallmark of the BK equation is the geometric scaling. It was found that the experimental data of the total cross section of the electron-proton deep inelastic scattering (DIS) at HEAR in small $x$ ($x<0.01$) region shows a geometric scaling behavior\cite{SGK}, which give a strong evidence of the CGC theory. However, a study based on the DGLAP evolution also shows geometric scaling behavior\cite{CF}. It is hard to distinguish which one (CGC or DGLAP) is the dominant mechanism to govern the evolution of the partonic system. To get more evidences to support the CGC mechanism, a lot of studies were carried out during the past years. On the one hand, a series of improved QCD evolution equations were proposed, such as running coupling BK (rcBK) equation\cite{Balitsky1,KW}, and the full next-to-leading-order (NLO) BK equation\cite{BC}. On the other hand, the CGC theory has been used to describe the experimental observables, like the proton structure function and the differential cross section for vector meson production, both from inclusive processes\cite{AM,CXY1,CXY2,YZhu1,YZhu2} and exclusive processes \cite{KMW,RS,GMM} which may offer more evidences for gluon saturation phenomenon.

In the field of CGC studies, the investigation of exclusive photoproduction of vector meson is especially important since it is very sensitive to the small $x$ gluons, thus it can offer an unique approach to probe the gluon saturation\cite{RSKV,GMS}. In particular, the quarkonia such as $J/\psi$ and $\phi$ are of great interest because they can explore not only the perturbative but also non-perturbative regimes. In recent years, these mesons have being investigated experimentally and theoretically. On the experimental side, the exclusive $J/\psi$ and $\phi$  photoproductions have been measured by H1 and ZEUS collaborations at HERA\cite{BBC,CDL,AMA,AAA2}. For higher energies, the LHCb collaboration at LHC has released the exclusive $J/\psi$ production data
in proton-proton (pp) collisions at $\sqrt{s}=7$ TeV and $\sqrt{s}=13$ TeV\cite{Aaij,Aaij2}, which enter even small $x$ region ($x\sim 10^{-6}$) and provide high precision experimental data to test the gluon saturation physics.

On the theoretical side, the pioneer study of gluon saturation by using diffractive DIS at HERA based on the Mueller's dipole model\cite{Mueller} can be traced back to two decades ago, in which the Geolec-Biernat and Wusthoff (GBW) model was firstly proposed to search the saturation effect\cite{GBW}. From then onwards, a lot of efforts were devoted to investigate the phenomenon of gluon saturation via diffraction in DIS. A dipole saturation model, which includes an impact parameter dependent, was developed to describe the differential diffractive $J/\psi$ production data at HERA\cite{KT,KMW}. It was shown that the $t-$distributions of differential cross sections are sensitive to saturation phenomena. An investigation in Ref.\cite{Munier04} showed that a good description of the diffractive DIS data was obtained by combing the dipole model with the Good and Walker picture. It was found that the diffractive observables can help to discriminate in an unique way between the predictions of different models in the saturation region. Based on the framework of the BK equation at non-zero momentum transfer, the authors in Ref.\cite{MPS} used the momentum transfer $\mathbf{q}$ instead of the impact parameter $\mathbf{b}$ in the saturation scale to devise an elegant formulism which is particularly convenient for comparisons between theoretical calculations and experimental data since the data are directly measured as a function of $t=-\mathbf{q}^2$. In order to investigate whether the diffractive photoproduction of vector meson is a sensitive probe of gluon saturation, a systematic study of the vector meson production was performed with two impact parameter dependent models\cite{AA}, the IP-Sat\cite{KT} and b-CGC\cite{WK}. The results further confirm that the $t$-distribution of differential cross sections of vector meson productions provides an unique method to discriminate among saturation and non-saturation models due to appearance of a pronounced dip in the $t$-distribution\cite{AA}. The aforementioned formulism was extended to study the vector meson productions in proton-proton and nucleus-nucleus collisions at LHC energies\cite{GMM}, which demonstrated that gluon saturation models can give a good description of the experimental data qualitatively. However, all the above mentioned saturation models for the vector meson production are based on the leading-order (LO) dipole amplitude which is inspired by the non-linear BK evolution equation at leading logarithmic accuracy in pQCD, which are insufficient for direct applications to phenomenology. It has been found that the evolution speed of the dipole amplitude resulting from the LO BK equation is too fast to give a precisely description of the HERA data, like proton structure functions\cite{IMMST,Levin16}.

Over the past decades, it has been shown that the higher order corrections have a significant contribution to the leading order BK equation. It has been shown that the LO evolution kernel is modified by the running coupling effect, which leads to the rcBK equation\cite{Balitsky1}. We would like to note that although kernel of the evolution equation is modified, the rcBK equation has the same structure as the LO BK equation. Our studies on the analytic solution to the rcBK equation has been demonstrated that the quadratic rapidity dependence in the exponent of the $S$-matrix in the LO case is replaced by the linear rapidity dependence once the running coupling correction is included, which indicate that the evolution speed of the dipole scattering amplitude is significantly suppressed by the running coupling effects\cite{Xiang1,Xiang2017}. The full NLO corrections, which includes quark loop (running coupling) and gluon loop contributions, to the evolution equation were calculated by Balitsky and Chirilli in Ref.\cite{BC}, they found that the kernel and structure of the evolution equation are changed by the full NLO effects. Note that the full NLO BK equation is unstable due to a large double transverse logarithm\cite{LM}, it can be stabilized by the resummation of the double logarithms leading to a collinearly-improved (ci) BK equation\cite{Iancu2015}. To see the influence of the full NLO corrections on the dipole scattering amplitude, we analytically solved the full NLO BK equation in the saturation region\cite{Xiang2}. Our result shows that the rapidity evolution of the dipole scattering amplitude is still suppressed by the full NLO effects, but the evolution speed is rebound as compared to the running coupling case due to a compensation effect made by gluon loops. Furthermore, our recent studies of the dependence of the dipole scattering amplitude on the running coupling prescriptions have been shown that the argument of the coupling has a great impact on the dipole amplitude\cite{Xiang2019}. We find that the rapidity evolution speed of the dipole amplitude is significantly slowed down by the smallest dipole size running coupling prescription.

Some of the above mentioned NLO theories of the high-energy scattering have been directly applied to phenomenology in the inclusive process. The authors in Refs.\cite{AAMS,AAMQS} used the dipole amplitude resulting from the rcBK equation to fit the inclusive small $x$ HERA data. They obtained a rather good description of the data since the running coupling effect significantly slows down the growth of the dipole amplitude with increasing energy. Soon after the collinearly-improved BK equation was established, several groups confronted the resummed equation with recent HERA data\cite{IMMST,Albacete2017,Cepila2019}, they found that the fit is rather successful and shows good stability up to virtualities as large as $Q^2=400 \mathrm{GeV}^2$ for the exchanged photon. Although the NLO corrections have been proved to be significant in the inclusive process, they are almost not applied to the vector meson production in the exclusive process in the literature. Based on the aforementioned applications of the NLO effects in inclusive process, we are confident to believe that the higher order corrections are also important in the diffractive vector meson production process. In this work, we find that the $\chi^2/d.o.f$ for elastic cross section (2.159) and differential cross section (2.056) in the leading logarithmic approximation are greatly improved after including one of the NLO corrections, running coupling (1.097 for the elastic cross section and 1.449 for differential cross section). This outcome indicates that the higher order corrections play an important role in the quantitative description of the diffractive vector meson production data. To ensure that the outcome is confident, we study the diffractive vector meson production with the dipole amplitude resulting from the collinearly-improved NLO BK equation. The results show that our theoretical calculations are good in agreement with the experimental measurement with $\chi^2/d.o.f$ 0.932 for the elastic cross section and 1.357 for differential cross section.

We extend our model which includes higher order corrections, to study the diffractive vector meson productions at LHC energies. We find that the model can give a rather good description of the $J/\psi$ data from 7 TeV and 13 TeV proton-proton peripheral collisions. The predictions of the diffractive $\phi$ production are provided with our model for 7 TeV and 13 TeV proton-proton peripheral collisions at LHC, see Table.\ref{table:5}. One can find that the model with LO dipole amplitude gives a larger total cross section, which would not be favored by the data, than the one obtained by NLO dipole amplitude. Again, we see that the high order corrections suppress the evolution of the dipole amplitude.


\section{Exclusive photoproduction of vector meson at NLO in the dipole formalism}
In this section we give a brief review of the formalism for exclusive vector meson photoproduction in dipole model. We firstly introduce the dipole model for calculation of the vector meson productions at non-zero momentum transfer in the CGC framework. We then present the evolution equations of the dipole amplitude which is a key ingredient in the dipole model. The vector meson wavefunctions, which are also a portion of the dipole model, are given in the last part of this section.


\subsection{Exclusive photoproduction of vector mesons in dipole model at non-zero momentum transfer}
\begin{figure}[b!]
\begin{center}
\includegraphics[width=0.6\textwidth]{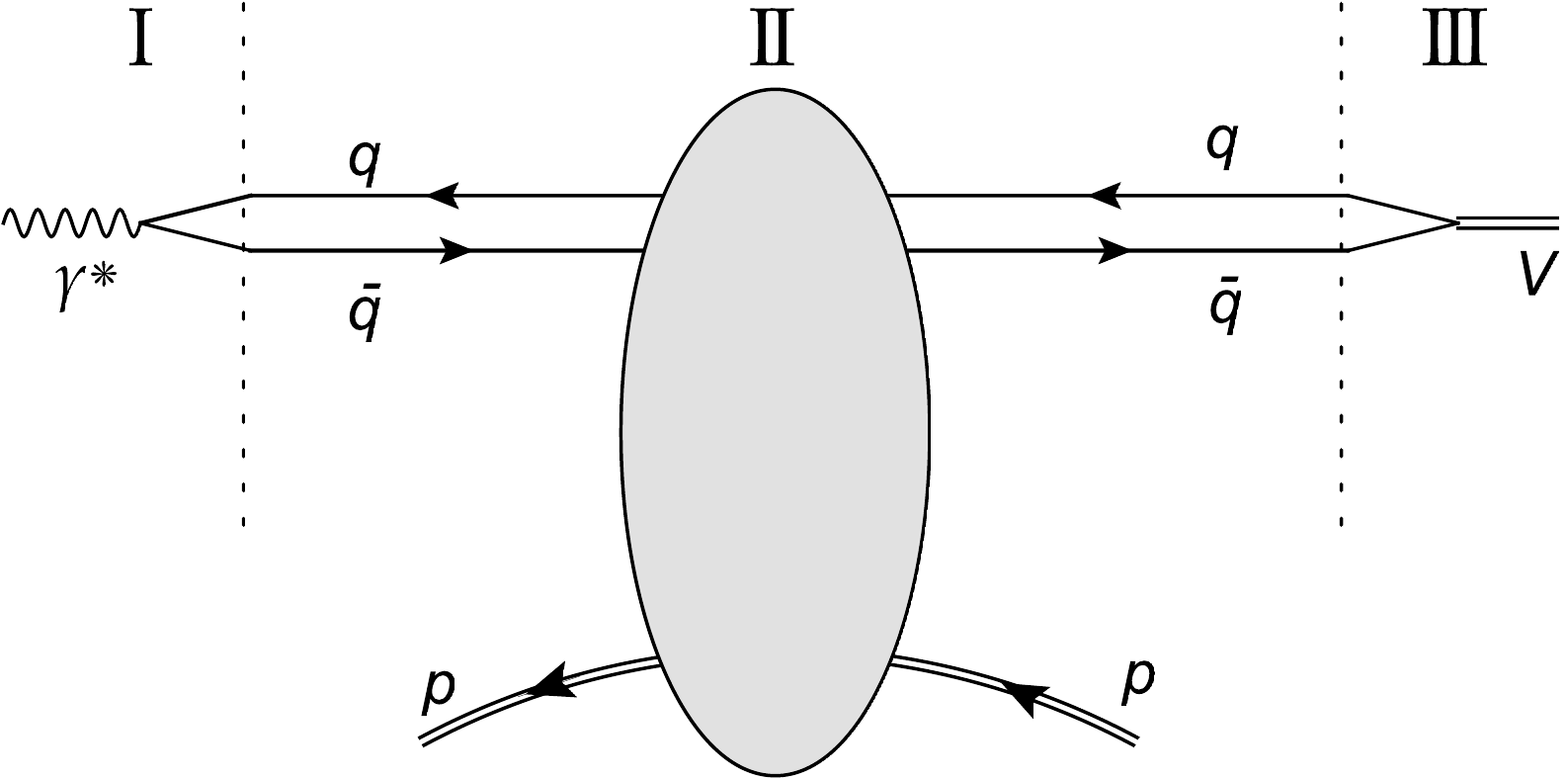}
\end{center}
\caption{The schematic diagram of a vector meson production in the $\gamma^* p\rightarrow Vp$ within the dipole model. The three separated subprocesses were denoted by $\text{\uppercase\expandafter{\romannumeral1}}$, $\text{\uppercase\expandafter{\romannumeral2}}$ and $\text{\uppercase\expandafter{\romannumeral3}}$, respectively.}
\label{dip}
\end{figure}
In terms of the dipole model, the vector meson production in an exclusive diffractive $\gamma^* p\rightarrow Vp$ scattering can be viewed as three separated subprocesses\cite{GM}, as shown in Fig.~\ref{dip}. The first subprocess is the formation of a dipole (a quark-antiquark pair) derived from a virtual photon fluctuation. The second subprocess is the interaction between the dipole and the proton via exchanging gluons. The last subprocess is the recombination of the outgoing quark-antiquark pair to produce a final vector meson. Therefore, the scattering amplitude of the diffractive process can be factorized into three ingredients: the photon wave function, the dipole-proton scattering amplitude and the vector meson wave function. Putting all ingredients together, one can write the imaginary part of the scatting amplitude for a vector meson production as
\be
\label{forward_A}
\mathcal{A}^{\gamma^* p\rightarrow Vp}_{T,L}(x,Q^2,\mathbf{q}) = \mathrm{i}\int_0^1\frac{dz}{4\pi}\int d^{2}\mathbf{r} \int d^{2}\mathbf{b}(\Psi_{V}^{*}\Psi)_{T,L}
 \mathrm{e}^{-\mathrm{i}\mathbf{b}\cdot\mathbf{q}} 2[1-S(x,\mathbf{r},\mathbf{b})],
\ee
where $z$ is the longitudinal momentum fraction of the incoming photon carried by quark, $x$ is the Bjorken variable, and $Q^2$ is the photon virtuality. The variable $\mathbf{q}$ denotes the momentum transfer, whose relationship with the squared momentum transfer is $t=-\mathbf{q}^2$. The remainder two dimensional vector $\mathbf{r}$ and $\mathbf{b}$ are the transverse size of the quark-antiquark dipole and the impact parameter, respectively. $\Psi$ is the wave function of the incoming photon, which can be accurately calculated by QED and is well known in the literature\cite{LB,DGKP}. $\Psi_{V}^{*}$ denotes the final vector meson wave function, unlike the photon wave function, it has various prescriptions as we shall discuss later at the end of this section. $(\Psi_{V}^{*}\Psi)_{T,L}$ represent the transverse and longitudinal overlap function between the photon and vector meson, respectively.

We would like to note that Eq.(\ref{forward_A}) is a scatting amplitude containing only the forward component. To get the nonforward scattering amplitude, one can multiply the forward wave functions by a phase factor $\exp[\pm\mathrm{i}(1-z)\mathbf{r}\cdot\mathbf{q}/2]$ as it was done in Ref.\cite{BGP}. Using this approach and assuming that the $S$-matrix is purely real (or the amplitude is purely imaginary), the scatting amplitude can be written as
\be
\label{nonforward_A}
\mathcal{A}^{\gamma^* p\rightarrow Vp}_{T,L}(x,Q^2,\mathbf{q}) = \mathrm{i}\int_0^1\frac{dz}{4\pi}\int d^{2}\mathbf{r} \int d^{2}\mathbf{b}(\Psi_{V}^{*}\Psi)_{T,L}
\mathrm{e}^{-\mathrm{i}[\mathbf{b}-(1-z)\mathbf{r}]\cdot\mathbf{q}} T(x,\mathbf{r},\mathbf{b}),
\ee
where $T(x,\mathbf{r},\mathbf{b})=1-S(x,\mathbf{r},\mathbf{b})$ describes the scattering amplitude between the dipole and proton, which contains all the basic information about the strong interactions between the dipole and proton. By taking into account the corrections from the real part of the scattering amplitude and the skewness effect, the differential cross section of an exclusive vector meson photoproduction can be written as\cite{KMW}:
\be
\label{dif_cross}
\frac{d\sigma^{\gamma^* p\rightarrow Vp}_{T,L}}{dt}
=\frac{(1+\beta^{2})R_{g}^{2}}{16\pi}\mid\mathcal{A}^{\gamma^* p\rightarrow Vp}_{T,L}(x,Q^2,\mathbf{q})\mid^{2},
\ee
\noindent where $\beta$ is the ratio of real part to imaginary part of the scattering amplitude and the factor $(1+\beta^{2})$ is to include the correction from the missing real part of the scatting amplitude due to the amplitude, $\mathcal{A}^{\gamma^* p\rightarrow Vp}_{T,L}$, in Eq.(\ref{nonforward_A}) only considering the contribution from the imaginary part. The skewness effect factor $R_{g}$ is derived from the fact that the momentum fraction of the exchanging gluons between the proton and dipole legs can be different. The parameters associated with these two corrections can be expressed by the imaginary part as follows:
\be
\label{corrections}
\beta = \tan(\frac{\pi\delta}{2}),~~\mathrm{and}~~
R_g= \frac{2^{2\delta+3}}{\sqrt{\pi}}\frac{\Gamma(\delta+5/2)}{\Gamma(\delta+4)},
\ee
with
\be
\label{delta}
\delta \equiv \frac{\partial\ln(\mathcal{A}_{T,L}^{\gamma^* p\rightarrow Vp})}{\partial\ln(1/x)}.
\ee

As we know that the dipole-proton scattering amplitude comes from the solution to the evolution equations, like the IIM model\cite{IIM} inspired by LO BK equation. In most cases, the impact parameter dependence is disregarded in the BK equation, since it was found that the dipole amplitude develops a power-like $b$ behaviour, called Coulomb tails, which render an unphysical results, i.e. the total cross section violation of the Froissart unitarity bound. To avoid the above mentioned difficulty, a general strategy is to build an impact parameter independent dipole amplitude inspired by the BK equation, then a model is employed to include the impact parameter dependence, such as two typical models IP-Sat\cite{KT} and b-CGC\cite{WK}. In this work, we use almost the same scheme as just mentioned but with impact parameter independent dipole amplitude resulting from numerical solution to the LO, rc, ci BK evolution equations. We introduce the impact parameter via multiplying the numerical dipole amplitude with a Gaussian $b$ dependence. In view of the advantage of the method\footnote{An elegant framework for calculation of $t$-distribution of vector meson productions is built based on the BK equation at non-zero momentum transfer, which is super convenient for comparison between theoretical calculations and experimental data due to the data directly measured as a function of $t$.}, which was proposed in Ref.\cite{MPS} by Marquet, Peschanski and Soyez (MPS), in study the $t$-distribution of differential cross sections of photoproduction of vector mesons, we shall follow the MPS strategy in this study.
Following Ref.\cite{MPS}, the dipole-proton scattering amplitude can be rewritten in terms of the momentum transfer $\mathbf{q}$ instead of the impact parameter $\mathbf{b}$ by using the Fourier transform
\be
\label{b2q}
\widetilde{T}(x,\mathbf{r},\mathbf{q}) =  \int d^{2}\mathbf{b}\mathrm{e}^{-\mathrm{i}\mathbf{b}\cdot\mathbf{q}} T(x,\mathbf{r},\mathbf{b}).
\ee
Substituting Eq.(\ref{b2q}) into Eq.(\ref{nonforward_A}), the scattering amplitude for $\gamma^* p\rightarrow Vp$ exclusive diffractive process becomes
\be
\label{A_q}
\mathcal{A}^{\gamma^* p\rightarrow Vp}_{T,L}(x,Q^2,\mathbf{q}) = \mathrm{i}\int_0^1\frac{dz}{4\pi}\int d^{2}\mathbf{r} (\Psi_{V}^{*}\Psi)_{T,L}
 \mathrm{e}^{\mathrm{i}z\mathbf{r}\cdot\mathbf{q}} \widetilde{T}(x,\mathbf{r},\mathbf{q}).
\ee
For the Fourier-transformed dipole-proton scattering amplitude $\widetilde{T}(x,\mathbf{r},\mathbf{q})$, we adopt a generalized formalism
\be
\label{T_q}
\widetilde{T}(x,\mathbf{r},\mathbf{q}) =  2\pi R^{2}\mathrm{e}^{-B\mathbf{q}^{2}} \mathcal{N}(r,x),
\ee
where the factor $\mathrm{e}^{-B\mathbf{q}^{2}}$ comes from the nonperturbative effects, $R$ can be interpreted as the radius of proton and $\mathcal{N}(r,x)$ is an impact parameter independent dipole amplitude. We would like to point out that $B$ and $R$ are free parameters in our fit, which shall be determined by fitting to HERA data.


\subsection{The dipole evolution equations}
A key ingredient to calculate the differential cross section is the dipole-proton scattering amplitude. It is known that almost all the past studies on the differential cross section of vector meson production in the framework of CGC were hovered on the LO level in the literature\cite{XC,GMN}. Although the LO dipole amplitude can describe the diffractive vector meson production experimental data at HERA under certain uncertainties\cite{FSS,MPS}, the precision of the model has to be improved to distinguish the dynamic mechanism of the CGC evolution from the DGLAP evolution, since the DGLAP formulism can also give a good description of the data\cite{AA}. Indeed a lot of efforts have been made to improve the accuracy of the CGC theory by including other higher order corrections, such as quark loops\cite{Balitsky1,KW}, gluon loops\cite{BC} and pomeron loops\cite{Xiang2007}. It was found that the running coupling effects dramatically slow down the evolution of the gluon system, which give a good description of the latest data from HERA on reduced cross sections\cite{AAMQS}. Similarly, the direct numerical solution of the full NLO BK equation is also shown that it slows down the evolution\cite{LM}. Based on the significance of the NLO corrections, we would like to extend LO vector meson production formalism to the NLO in this work. One shall see in the next section that the descriptions of the experimental data are dramatically improved once the NLO corrections are included.

The LO BK equation describes the evolution of a quark-antiquark (with a quark at $x_{\bot}$ and an antiquark at $y_{\bot}$) dipole with the rapidity $Y$ by the emission of a soft gluon. In large $N_c$ limit, it can be written as
\be
\label{BK}
\frac{\partial N(r,Y)}{\partial Y}  =
         \int d^2z_{\bot} K^{\mathrm{LO}}
           \left [ N(r_1,Y) + N(r_2,Y) -N(r,Y)- N(r_1,Y)N(r_2,Y) \right ] ,
\ee
with the evolution kernel
\be
\label{BK_ker}
K^{\mathrm{LO}}=\frac{\bar{\alpha}_s}{2 \pi} \frac{r^2}{r_1^2r_2^2},
\ee
where $\bar{\alpha}_s = \alpha_sN_c/\pi$. Here $z_{\bot}$ denotes the transverse coordinate of emitted gluon in the evolution. In Eq.(\ref{BK}) we have used the notation ${\bf r}=x_{\bot}-y_{\bot}$, ${\bf r}_1=x_{\bot}-z_{\bot}$ and ${\bf r}_2=z_{\bot}-y_{\bot}$ to denote the transverse size of parent and the new daughter dipoles, respectively. The BK equation is obtained at leading logarithmic approximation, it has been found that it is insufficient when confronting with experimental data\cite{AAMS,AAMQS,AAMSW}. Therefore, a lot of efforts have been made to improve the understanding of the dipole's evolution at NLO accurary.

The first improvement to the LO BK equation was done by including quark loops. After resumming $\alpha_s N_f$ to all orders, one can get an evolution equation with running coupling corrections\cite{Balitsky1,KW}, which is called as rcBK equation. The rcBK equation is given by
\be
\label{rcBK}
\frac{\partial N(r,Y)}{\partial Y} =
\int d^2z_{\bot}K^{\mathrm{rc}} \left[ N(r_1,Y) + N(r_2,Y)-N(r,Y)- N(r_1,Y) N(r_2,Y) \right] ,
\ee
with a modified evolution kernel
\be
\label{rcBK_ker}
K^{\mathrm{rc}}=\frac{\bar{\alpha}_s}{2 \pi}
  \left[\frac{r^2}{r_1^2\,r_2^2}+
    \frac{1}{r_1^2}\left(\frac{\alpha_s(r_1^2)}{\alpha_s(r_2^2)}-1\right)+
    \frac{1}{r_2^2}\left(\frac{\alpha_s(r_2^2)}{\alpha_s(r_1^2)}-1\right)
  \right].
\ee
The numerical solution of the rcBK equation has been obtained by Albacete et al.\cite{AAMS,AAMQS}, they found the proton structure function can be well described under this evolution equation. However, the quark loops corrections are not the only source of the higher order corrections, the complete NLO corrections should also include the contributions from gluon loops and the tree gluon diagrams with quadratic and cubic nonlinearities\cite{BC}. Considering all these contributions, one can get the full NLO BK evolution equation
 \bea
 \frac{\partial N(r, Y)}{\partial Y} & = & \frac{\bar{\alpha}_s}{2\pi}\int d^2r_1K_1\left[ N(r_1,Y) + N(r_2,Y)-N(r,Y)- N(r_1,Y)N(r_2,Y) \right]
 + \frac{\bar{\alpha}_s^2}{8\pi^2}\int d^2r_1d^2r_2'K_2
 \nonumber\\
&& \times \left[ N(r_3,Y) + N(r_2',Y) + N(r_1,Y) N(r_2,Y) + N(r_1,Y)N(r_3,Y)N(r_2',Y)-N(r_2,Y)\right.\nonumber\\
&& \left.- N(r_1,Y)N(r_3,Y)- N(r_1,Y)N(r_2',Y) - N(r_3,Y)N(r_2',Y)\right] + \frac{\bar{\alpha}_s^2N_f}{8\pi^2N_c}\int d^2r_1d^2r_2' K_3 \nonumber\\
&& \times \left[ N(r_1',Y) + N(r_1,Y)N(r_2,Y) - N(r_1,Y) - N(r_1',Y)N(r_2,Y) \right],
 \label{fnlobk0}
 \eea

 where the kernels are
 \bea
K_1 &=& \frac{r^2}{r_1^2r_2^2} + \frac{1}{r_1^2}\bigg(\frac{\alpha_s(r_1^2)}{\alpha_s(r_2^2)}-1\bigg)+\frac{1}{r_2^2}\bigg(\frac{\alpha_s(r_2^2)}{\alpha_s(r_1^2)}-1\bigg)
 +\frac{\bar{\alpha}_s(r^2)r^2}{r_1^2r_2^2}\bigg(\frac{67}{36} - \frac{\pi^2}{12} - \frac{5N_f}{18N_c} - \frac{1}{2}\ln\frac{r_1^2}{r^2}\ln\frac{r_2^2}{r^2}\bigg), \label{fnlobk1k1} \\
K_2 &=& -\frac{2}{r_3^4} + \bigg[\frac{r_1^2r_2'^2 + r_1'^2r_2^2 - 4r^2r_3^2}{r_3^4(r_1^2r_2'^2-r_1'^2r_2^2)} + \frac{r^4}{r_1^2r_2'^2(r_1^2r_2'^2 - r_1'^2r_2^2)} +
\frac{r^2}{r_1^2r_2'^2r_3^2}\bigg]\ln\frac{r_1^2r_2'^2}{r_1'^2r_2^2}, \label{fnlobk1k2}\\
K_3 &=& \frac{2}{r_3^4} - \frac{r_1'^2r_2^2 + r_2'^2r_1^2 - r^2r_3^2}{r_3^4(r_1^2r_2'^2 - r_1'^2r_2^2)}\ln\frac{r_1^2r_2'^2}{r_1'^2r_2^2}. \label{fnlobk1k3}
\eea
In Eq.(\ref{fnlobk0}) we have used the notation ${\bf r}_1'=x_{\bot}-z_{\bot}'$, ${\bf r}_2'=y_{\bot}-z_{\bot}'$, and ${\bf r}_3=z_{\bot}-z_{\bot}'$ to denote the transverse size of dipoles.

From Eq.(\ref{fnlobk1k1}) we can see there is a double logarithmic term $\ln\frac{r_1^2}{r^2}\ln\frac{r_2^2}{r^2}$ in the evolution kernel, which renders the full NLO BK equation unstable\cite{LM}. The solution can turn to a negative value for some region due to the double logarithmic term. So, one needs to make a resummation of these double logarithms under the double logarithmic approximation (DLA) as it has done by Iancu et al. in Ref.\cite{Iancu2015}. When this resummation is applied to the full NLO BK equation, the double logarithmic term is removed from kernel $K_1$, and the resummation will modify kernel $K_1$ by multiplying it with kernel
\be
\label{DLA_ker}
K^{\mathrm{DLA}}=\frac{J_1\bigg(2\sqrt{\abar \rho^2}\bigg)}{\sqrt{\abar\rho^2}}\simeq 1 - \frac{\abar\rho^2}{2} + \mathcal{O}(\abar^2),
\ee
with $\rho=\sqrt{\ln\frac{r_1^2}{r^2}\ln\frac{r_2^2}{r^2}}$.

In addition to the double logarithmic term, the single transverse logarithms (STL) will also generate a large logarithmic corrections to the evolution equation as shown in Ref.\cite{IMMST}. The effect of the single transverse logarithm resummation will also modify kernel $K_1$ by multiplying it with kernel
\be
K^{\mathrm{STL}} = \exp\bigg\{-\abar A_1\bigg|\ln\frac{r^2}{\mathrm{min}\{r_1^2, r_2^2\}}\bigg|\bigg\}.
\ee
with anomalous dimension $A_1=\frac{11}{12}$.

By resumming the large single and double transverse logarithms as in Ref.\cite{IMMST}, the collinearly-improved version of BK evolution equation reads
 \bea
 \frac{\partial N(r, Y)}{\partial Y} & = & \frac{\bar{\alpha}_s}{2\pi}\int d^2r_1K_1^{\mathrm{CI}}\left[ N(r_1,Y) + N(r_2,Y)-N(r,Y)- N(r_1,Y)N(r_2,Y) \right]
 + \frac{\bar{\alpha}_s^2}{8\pi^2}\int d^2r_1d^2r_2'K_2
 \nonumber\\
&& \times \left[ N(r_3,Y) + N(r_2',Y) + N(r_1,Y) N(r_2,Y) + N(r_1,Y)N(r_3,Y)N(r_2',Y)-N(r_2,Y)\right.\nonumber\\
&& \left.- N(r_1,Y)N(r_3,Y)- N(r_1,Y)N(r_2',Y) - N(r_3,Y)N(r_2',Y)\right] + \frac{\bar{\alpha}_s^2N_f}{8\pi^2N_c}\int d^2r_1d^2r_2' K_3 \nonumber\\
&& \times \left[ N(r_1',Y) + N(r_1,Y)N(r_2,Y) - N(r_1,Y) - N(r_1',Y)N(r_2,Y) \right],
 \label{fnlobk1}
 \eea
where the collinearly improved kernel in the first integration term becomes
\bea
K_1^{\mathrm{CI}} &=& K^{\mathrm{DLA}}K^{\mathrm{STL}}\bigg[\frac{r^2}{r_1^2r_2^2} + \frac{1}{r_1^2}\bigg(\frac{\alpha_s(r_1^2)}{\alpha_s(r_2^2)}-1\bigg)+\frac{1}{r_2^2}\bigg(\frac{\alpha_s(r_2^2)}{\alpha_s(r_1^2)}-1\bigg)\bigg]\nonumber\\
&& - \frac{r^2}{r_1^2r_2^2}\bigg(-\abar A_1\bigg|\ln\frac{r^2}{\mathrm{min}\{r_1^2, r_2^2\}}\bigg|\bigg)
+\frac{\abar(r^2)r^2}{r_1^2r_2^2}\bigg(\frac{67}{36} - \frac{\pi^2}{12} - \frac{5N_f}{18N_c}\bigg).
\label{ci_kernel}
\eea
Note that the Eqs.(\ref{BK}), (\ref{rcBK}) and (\ref{fnlobk1}) shall be numerically solved, and their solutions shall be used as dipole amplitudes to calculate the elastic and differential cross sections in the next section.


\subsection{The wavefunctions for vector meson}
Another ingredient to compute the differential cross section for vector meson production is the overlap function $(\Psi_{V}^{*}\Psi)_{T,L}$, which depends on the quark momentum fraction $z$, the dipole transverse size $\mathbf{r}$ and the photon virtuality $Q^{2}$. The overlap function has various prescriptions, such as boosted Gaussian, Gauss-LC and DGKP\cite{KMW}. Although it has been shown in Ref.\cite{MPS} that for an identified meson not all the overlap functions can give equally good description of the experimental data, namely a meson has its own favorite wavefunction. We focus on studying the higher order effects for vector meson production in this study. So, we shall use an unified formalism of wavefunction for different mesons to have a better insight into the higher order effects. The overlap function between the photon and the vector meson has transverse and longitudinal components and can be written as\cite{KMW}
\be
\label{eq:overlapT}
  (\Psi_V^*\Psi)_{T} = \hat{e}_f e \frac{N_c}{\pi z(1-z)}\Big\{m_f^2 K_0(\epsilon r)\phi_T(r,z)
- [z^2+(1-z)^2]\epsilon K_1(\epsilon r) \partial_r \phi_T(r,z)\Big\},
\ee
\be
\label{eq:overlapL}
  (\Psi_V^*\Psi)_{L} = \hat{e}_f e \frac{N_c}{\pi}2Qz(1-z)K_0(\epsilon r)
\Big[M_V\phi_L(r,z)+ \delta\frac{m_f^2 - \nabla_r^2}{M_Vz(1-z)}
    \phi_L(r,z)\Big].
\ee
In Eqs.(\ref{eq:overlapT}) and (\ref{eq:overlapL}), $\phi(r,z)$ is the scalar function. In our study the boosted Gaussian scalar functions are used since it works well for both light and heavy mesons\cite{CFS}. In boosted Gaussian formalism, the scalar functions are given by
\be
\label{eq:BGT}
\phi_{T}(r,z) = \mathcal{N}_{T} z(1-z)\exp\Big(-\frac{m_f^2 \mathcal{R}_{T}^2}{8z(1-z)}
- \frac{2z(1-z)r^2}{\mathcal{R}_{T}^2} + \frac{m_f^2\mathcal{R}_{T}^2}{2}\Big),
\ee
\be
\label{eq:BGL}
\phi_{L}(r,z) = \mathcal{N}_{L} z(1-z)\exp\Big(-\frac{m_f^2 \mathcal{R}_{L}^2}{8z(1-z)}
- \frac{2z(1-z)r^2}{\mathcal{R}_{L}^2} + \frac{m_f^2\mathcal{R}_{L}^2}{2}\Big).
\ee
%
%
%
The variable $\epsilon$ in the Bessel functions in Eqs.(\ref{eq:overlapT}) and (\ref{eq:overlapL}) is $\epsilon^{2}=z(1-z)Q^{2}+m_f^2$. The values of the parameters $M_V$, $m_f$, $N_{T,L}$, and $R_{T,L}$ in the above equations are given in Table~\ref{table:1}. It is worth to note that the longitudinal component is ignored in most studies due to its small contribution\cite{GMN, XC2}. It is safe in a very small photon virtuality regime, like quasi-real photoproduction. However, the longitudinal component can give a significant contribution in a large photon virtuality region as we have discussed in Ref.\cite{CYZX}. So, the longitudinal component is included in this study since we confront with data at variety photon virtualities.
\begin{table}[htbp]
  \begin{center}
  \begin{tabular}{cc|cccccc}
  \hline
  &Meson &\quad $M_V/\text{GeV}$ &\quad $m_f/\text{GeV}$ &\quad $N_{T}$ &\quad $N_{L}$ &\quad $R_{T}/\text{GeV}^{-2}$ &\quad $R_{L}/\text{GeV}^{-2}$ \\
  \hline
  &  $J/\psi$  &\quad 3.097    &\quad 1.4    &\quad 0.578    &\quad  0.575    &\quad 2.3    &\quad 2.3 \\
  \hline
  &  $\phi$ &\quad 1.019   &\quad 0.14    &\quad 0.919    &\quad  0.825    &\quad  11.2    &\quad 11.2 \\
  \hline
    \end{tabular}%
  \caption{Parameters of the boosted Gaussian formalism for $J/\psi$ and $\phi$\cite{KMW}.}
  \label{table:1}
  \end{center}
\end{table}%



\section{Numerical Results}
In this section, we use the dipole amplitudes, which come from the numerical solutions to the LO, rc and ci BK evolution equations, to calculate the vector meson productions. Firstly, we give a brief description about numerical method to solve differential equations and experimental data sets used in our fit. Then, we show our theoretical calculations of $J/\psi$ and $\phi$ productions and compare them with the experimental data from HERA. Finally, we extend the formalism to LHC energies and make predictions for the rapidity distributions of $J/\psi$ and $\phi$ productions in pp collisions at 7 TeV and 13 TeV.


\subsection{Numerical setup and data selection}
As well known that the LO, rc and ci BK evolution equations are integro-differential equations. To get the numerical solutions, we can solve them on a lattice. In this study, we discretize the variable $r$ into 256 points ($r_{min}=2.06 \times 10^{-9}$ and $r_{max}=54.6$). Throughout this numerical study, the unit of dipole size $r$ is $\mathrm{GeV}^{-1}$. For the rapidity, the number of points are set to 100 with the step size $\triangle Y=0.2$. This setup can insure that the grid is small enough for our purpose. To perform the numerical simulations, we employ the GNU scientific library (GSL). The main GSL subroutines we have used are Runge-Kutta for solving ordinary differential equations, the adaptive integration for numerical integrals, and the cubic spline interpolation for interpolating data points.

To solve these integro-differential equations, one needs initial conditions. There are several kinds of the initial conditions in the literature, such as GBW\cite{GBW} and MV\cite{MV} models. The Refs.\cite{AAMS} and \cite{AAMQS} are used both of them as initial conditions for the rcBK equation in the fit of the reduced cross sections, they showed that the MV initial condition is much more favorable by the experimental data than the GBW model. So, we adopt the MV model as initial condition in this work\cite{MV},
\be
\label{eq:MV}
{N}(r,Y\!=\!0)=1-\exp{\left[-\left(\frac{r^2Q_{s0}^{2}}{4}\right)^{\gamma}
    \log{\left(\frac{1}{r\,\Lambda_{QCD}}+e\right)}\right]},
\ee
with $\gamma=1.13$, $Q^2_{s0}=0.15\, \mathrm{GeV^2}$, and $\Lambda_{QCD}=0.241\, \mathrm{GeV}$.

%

In our analysis, we use Eq.(\ref{dif_cross}) to fit the differential cross section and the elastic cross section for $J/\psi$ and $\phi$ productions. The experimental data are taken from ZEUS Collaboration ($J/\psi$\cite{CDL}, $\phi$\cite{BBC}) and H1 Collaboration ($J/\psi$\cite{AAA2}, $\phi$\cite{AMA}). It should be noted that our studies are within the framework of the CGC which is valid in the range $x \leq x_{0}$ with $x_0=10^{-2}$. Therefore, the data points with $x$ larger than $x_{0}$ are automatically excluded in the data set. In addition, we exclude the data with large error bars at large photon virtuality. After selection, the total number of data are 177 points, which are used in the fit. For the details, the elastic cross section data of $J/\psi$ and $\phi$ productions are 58 and 61 points, and the differential cross section data of $J/\psi$ and $\phi$ productions are 24 and 34 points, respectively.
\begin{table}[htbp]
  \begin{center}
  \begin{tabular}{cc|cccccc}
  \hline
  &  $\mathcal{N}(r,x)$ & $B/\text{GeV}^{-2}$  &$R/\text{GeV}^{-1}$ &$\chi^{2}/d.o.f$ \\
  \hline
  & LO    & 2.500   & 3.800   & 2.159     \\
  \hline
  &  rc  & 1.954    & 3.791    & 1.097     \\
  \hline
  &  ci & 2.060    & 3.737    & 0.932     \\
  \hline
    \end{tabular}%
  \caption{Parameters and $\chi^{2}/d.o.f$ results for the elastic cross section with different dipole amplitudes.}
  \label{table:2}
  \end{center}
\end{table}%
\begin{table}[htbp]
  \begin{center}
  \begin{tabular}{cc|cccccc}
  \hline
  &  $\mathcal{N}(r,x)$ & $B/\text{GeV}^{-2}$  &$R/\text{GeV}^{-1}$ &$\chi^{2}/d.o.f$ \\
  \hline
  & LO    & 2.253    & 3.223   & 2.056     \\
  \hline
  &  rc  & 2.200    & 3.480    & 1.449     \\
  \hline
  &  ci & 2.175    & 3.349    & 1.357     \\
  \hline
    \end{tabular}%
  \caption{Parameters and $\chi^{2}/d.o.f$ results for the differential cross section with different dipole amplitudes.}
  \label{table:3}
  \end{center}
\end{table}%


\subsection{Fitting results with HERA data}
To demonstrate the significance of the high order corrections in the description of the HERA data, one needs to compute the vector meson productions with the LO and NLO dipole amplitudes and compare these calculations. In this study, there are only two free parameters $B$ and $R$, see Eq.(\ref{T_q}), other parameters like, $x_0$, $C^{2}$, are directly taken from\cite{AAMS}. Tables~\ref{table:2} and \ref{table:3} show these two parameters and $\chi^{2}/d.o.f$ results from our fit. From the values of the $\chi^{2}/d.o.f$ in the last columns in Tables~\ref{table:2} and \ref{table:3}, one can see that the NLO descriptions of the vector meson productions are better than the LO case, which indicate that the NLO corrections play an important role in diffractive process. Especially, the $\chi^{2}/d.o.f$ resulting from the fit to the elastic cross section, $\sigma$, there is a large improvement as compared to the LO description once the NLO corrections are included. By Global analysis, one can see that the values of $\chi^{2}/d.o.f$ calculated from the rc and ci dipole amplitudes are more close to unit than those calculated from the LO amplitude.
\begin{figure}[h!]
\begin{center}
\includegraphics[width=0.48\textwidth]{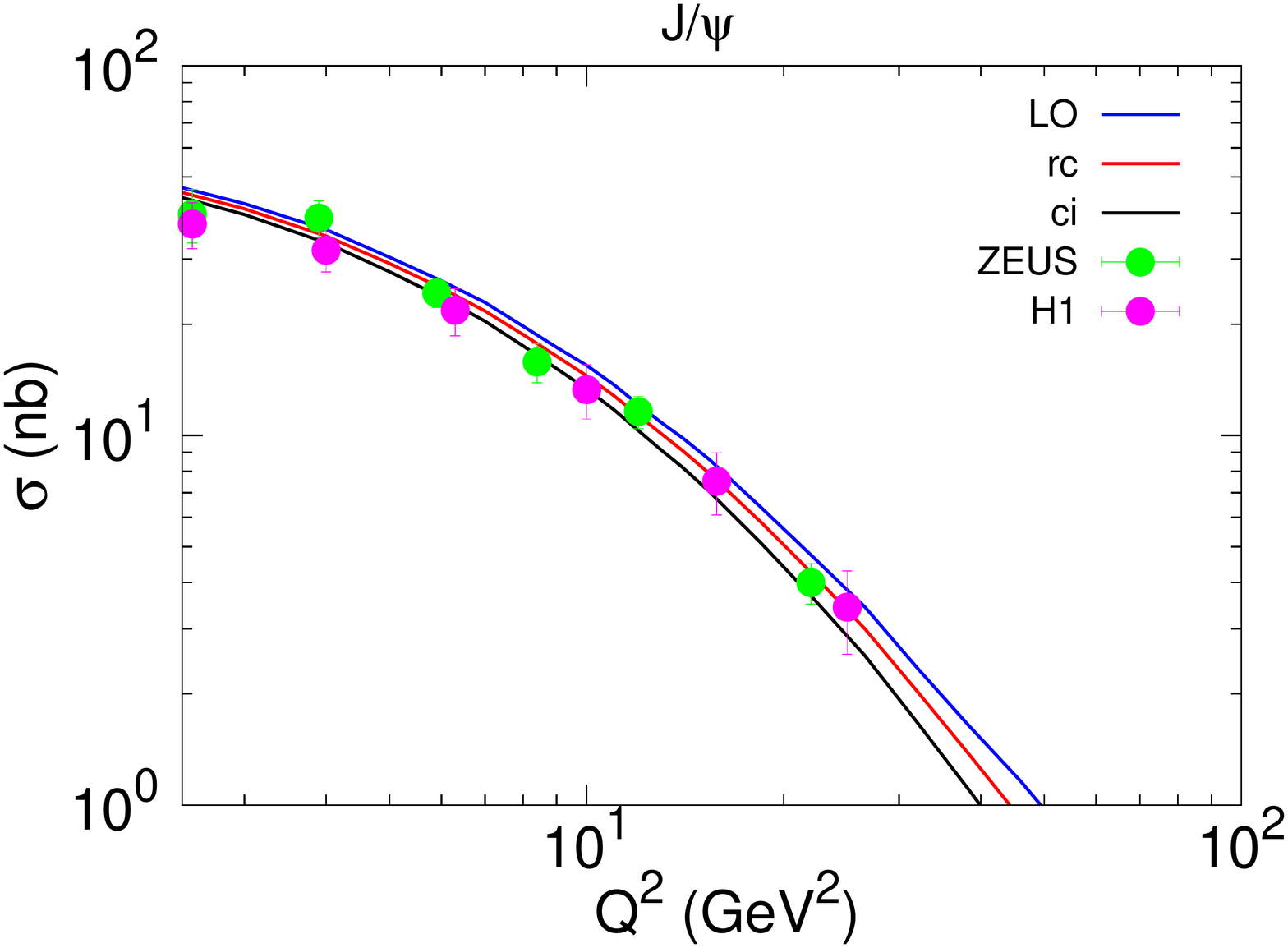}
\vspace{0cm}\hspace{-0.3cm}
\includegraphics[width=0.48\textwidth]{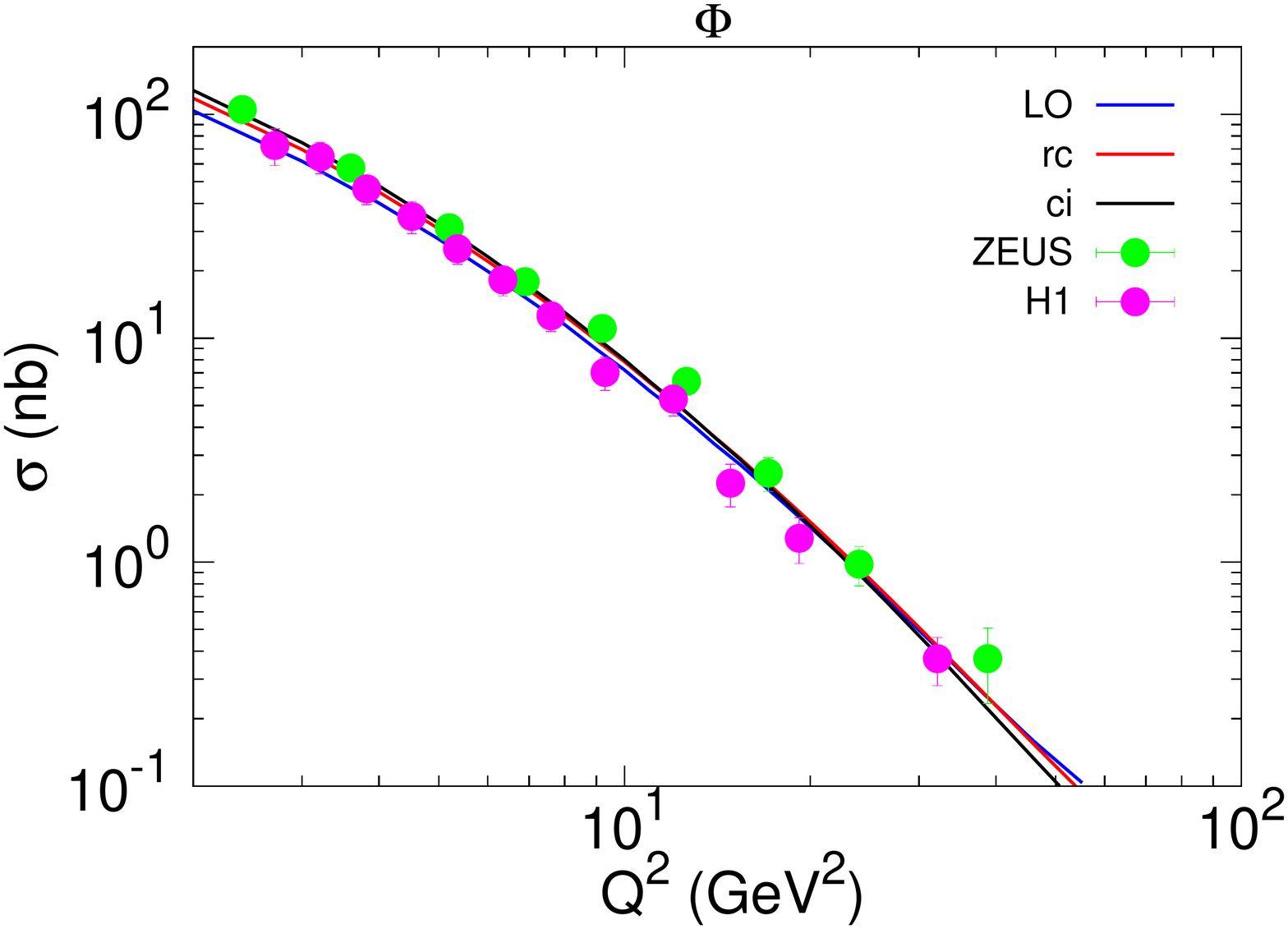}
\end{center}
\vspace{-0.5cm}
\caption{Elastic cross section $\sigma$ for $J/\psi$ and $\phi$ as a function of $Q^{2}$.}
\label{Q2}
\end{figure}
\begin{figure}[h!]
\begin{center}
\includegraphics[width=0.48\textwidth]{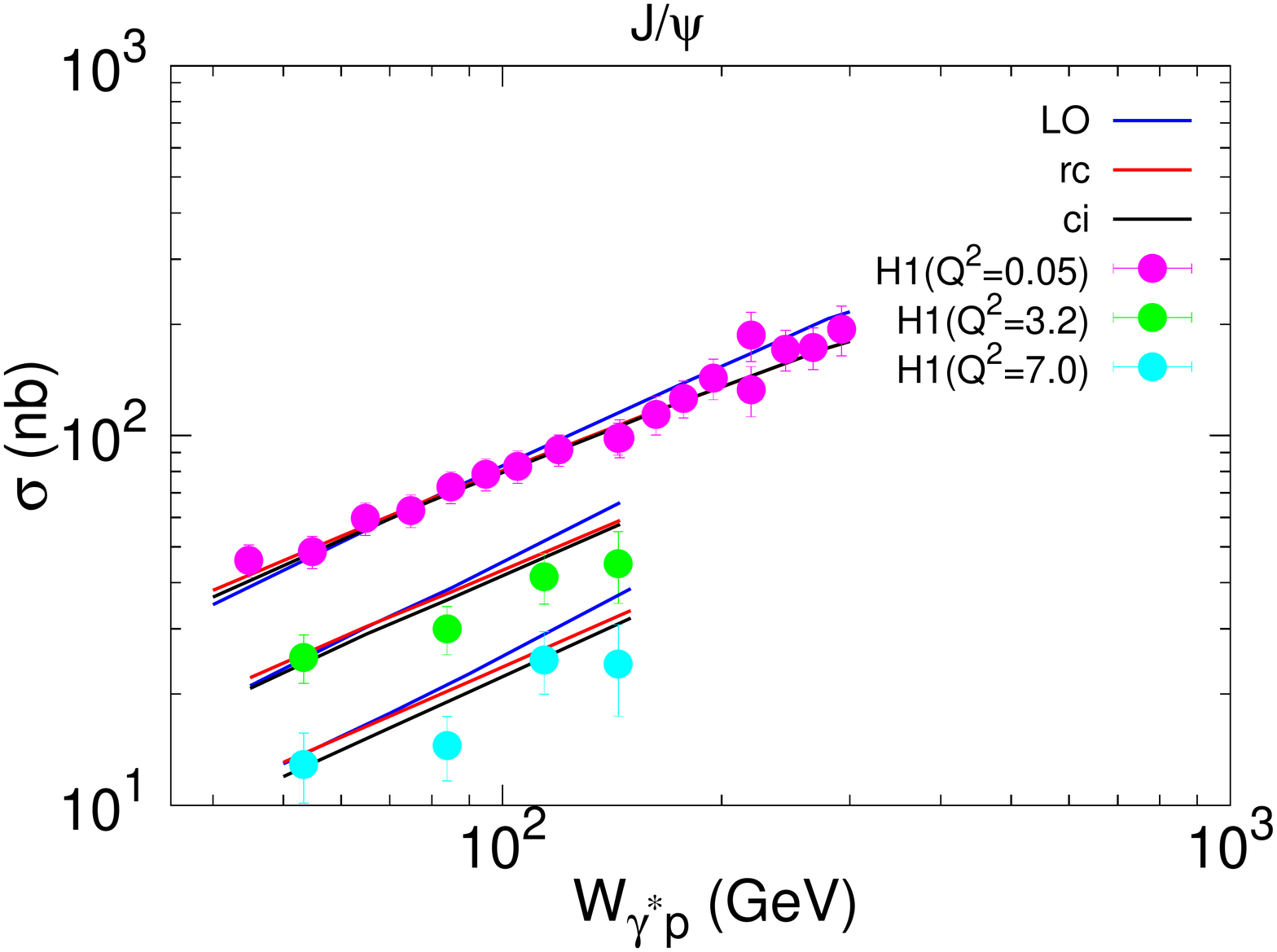}
\vspace{0cm}\hspace{-0.3cm}
\includegraphics[width=0.48\textwidth]{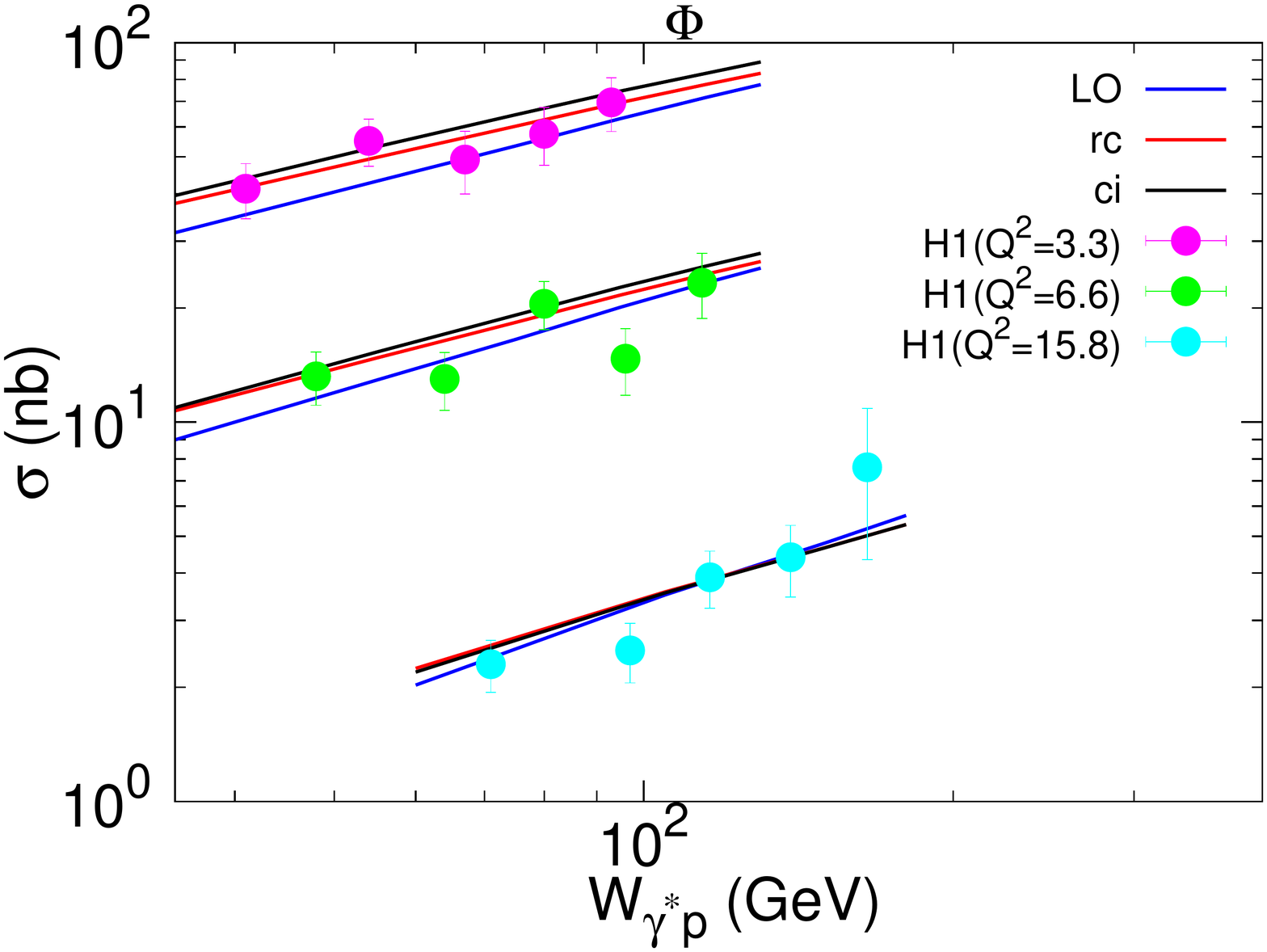}
\vspace{0cm}\hspace{0cm}
\includegraphics[width=0.48\textwidth]{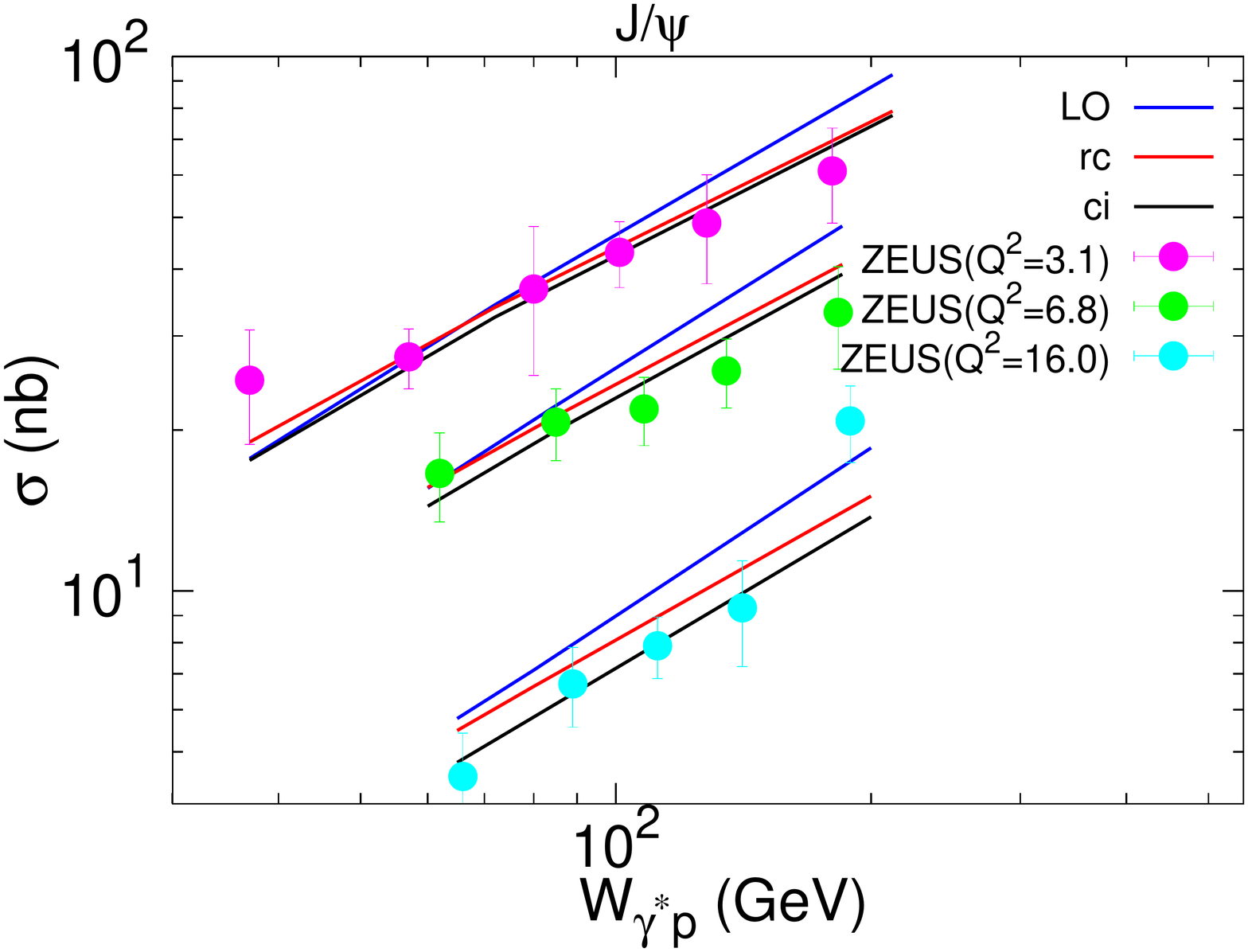}
\vspace{0cm}\hspace{-0.3cm}
\includegraphics[width=0.48\textwidth]{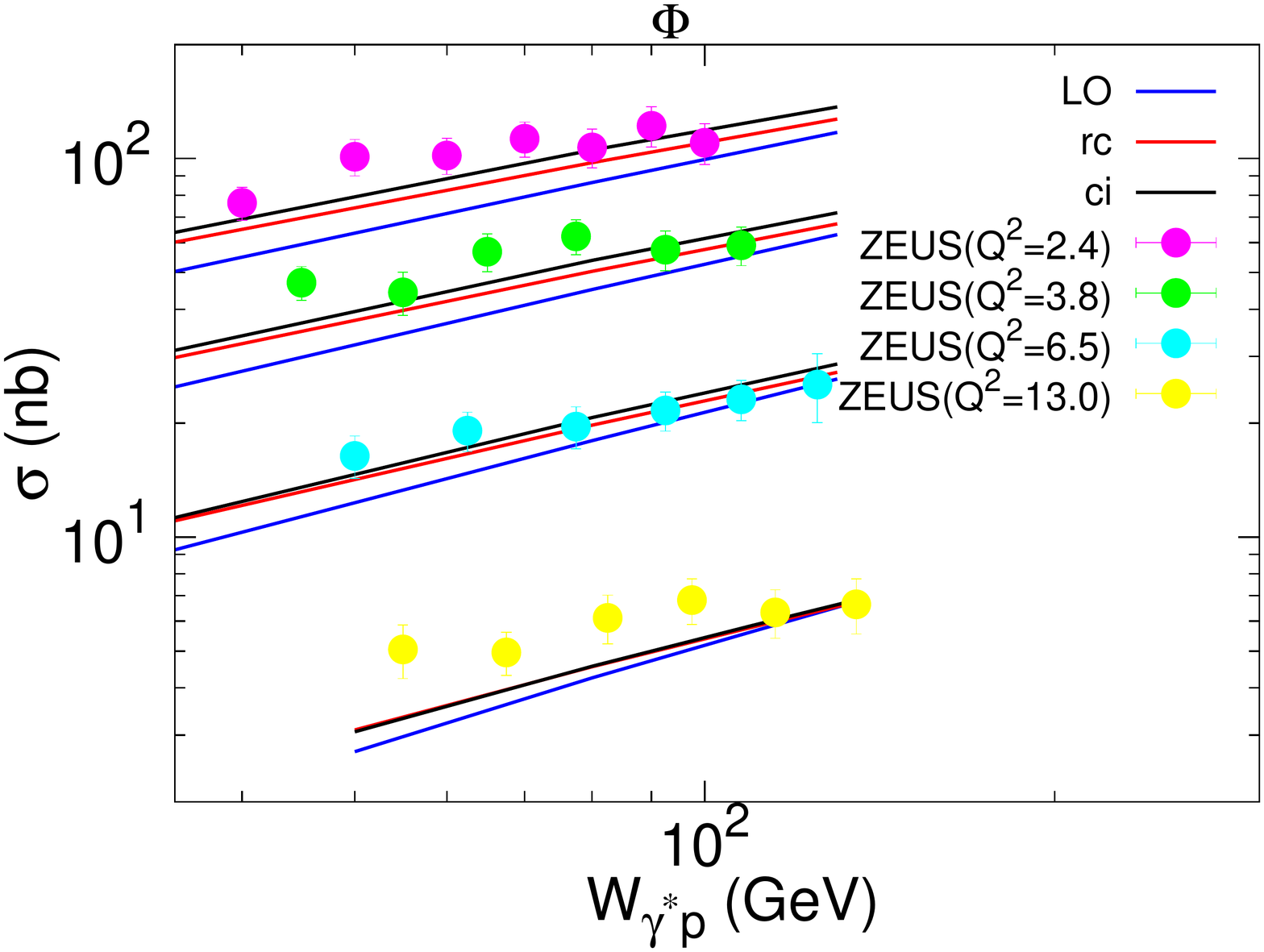}
\end{center}
\vspace{-0.5cm}
\caption{Elastic cross section $\sigma$ for $J/\psi$ and $\phi$ as a function of $W_\gamma p$ at different $Q^{2}$.}
\label{W}
\end{figure}
\begin{figure}[h!]
\setlength{\unitlength}{1.5cm}
\begin{center}
\includegraphics[width=0.48\textwidth]{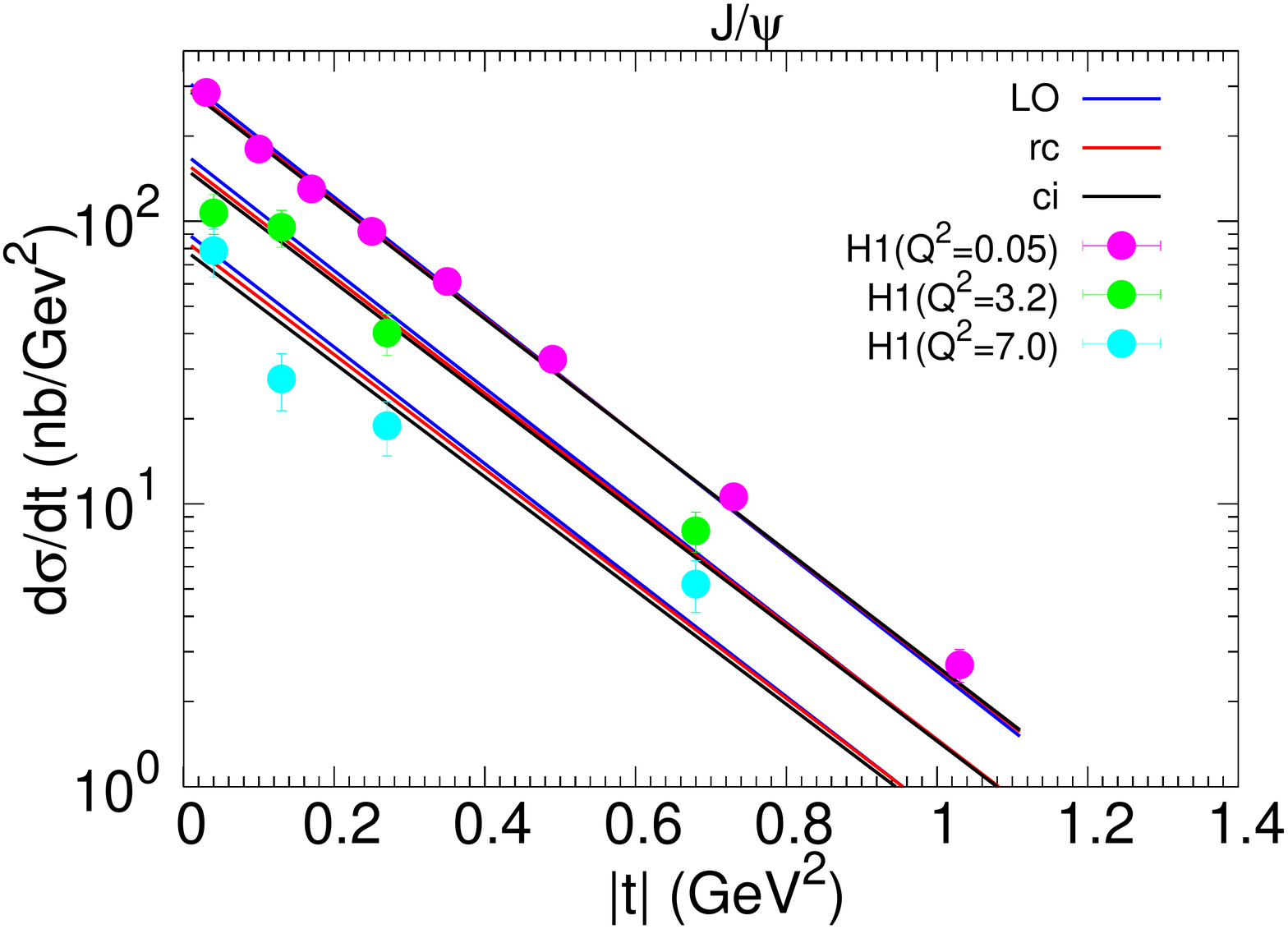}
\vspace{0cm}\hspace{-0.3cm}
\includegraphics[width=0.48\textwidth]{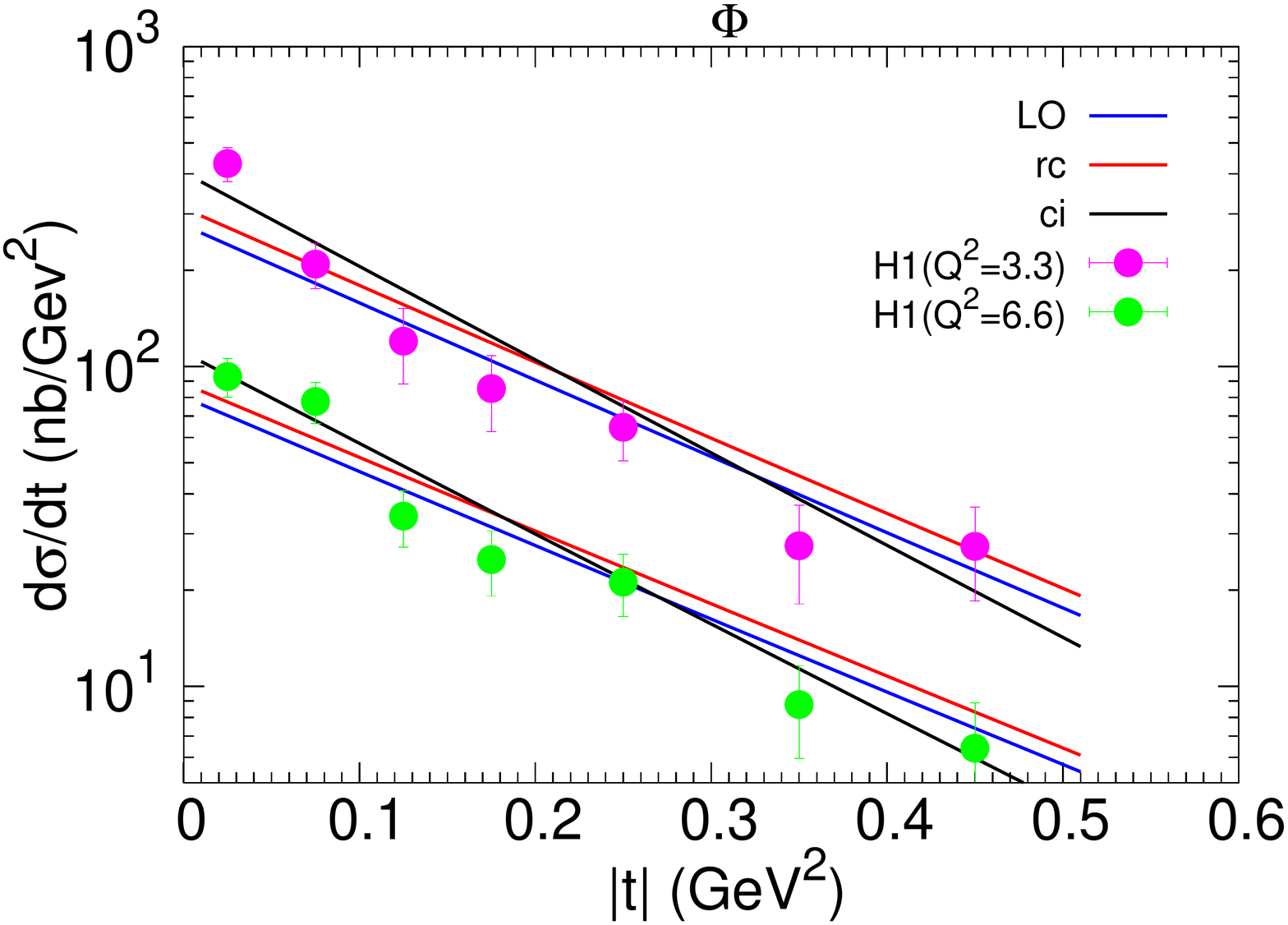}
\vspace{0cm}\hspace{0cm}
\includegraphics[width=0.48\textwidth]{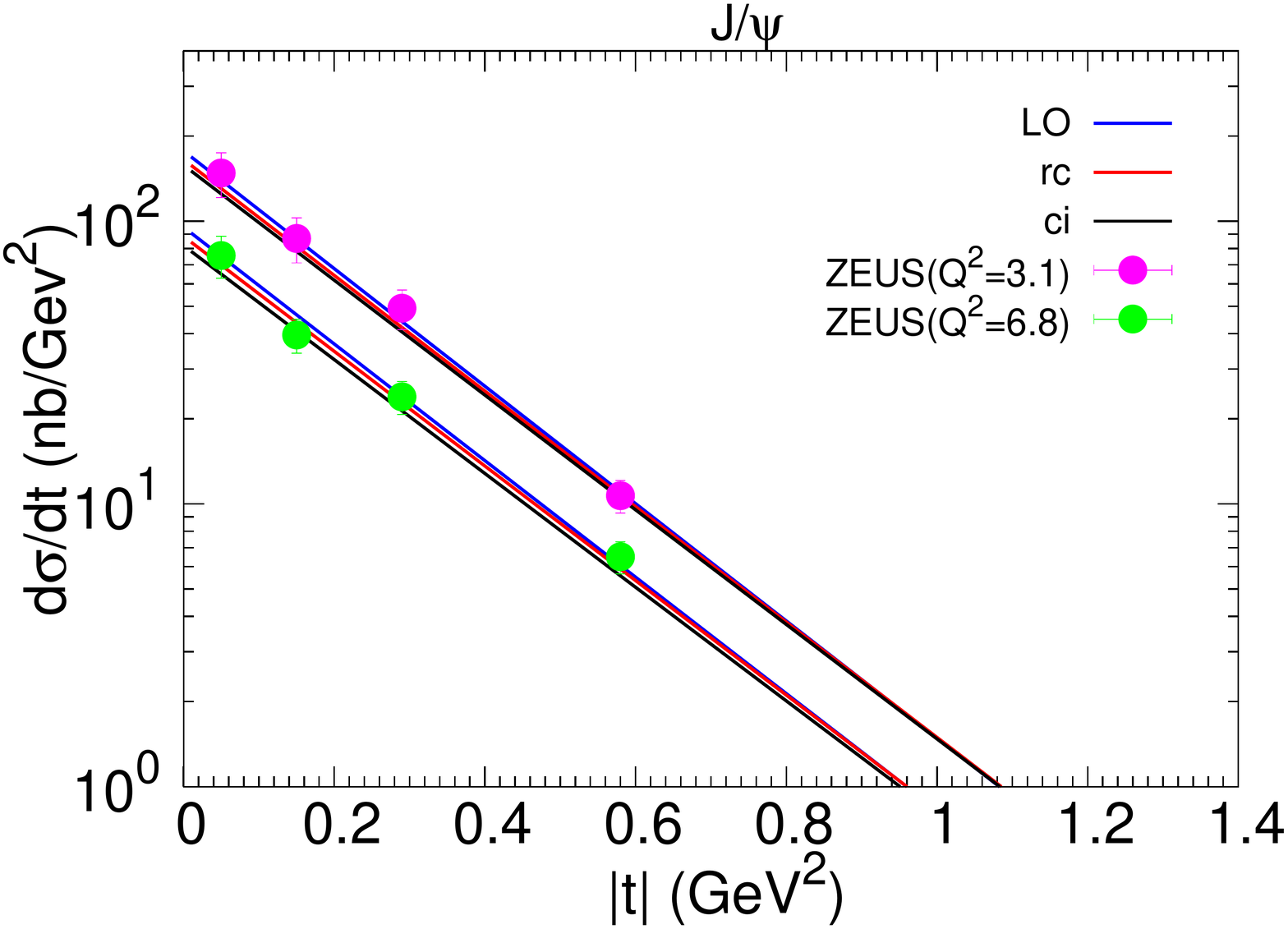}
\vspace{0cm}\hspace{-0.3cm}
\includegraphics[width=0.48\textwidth]{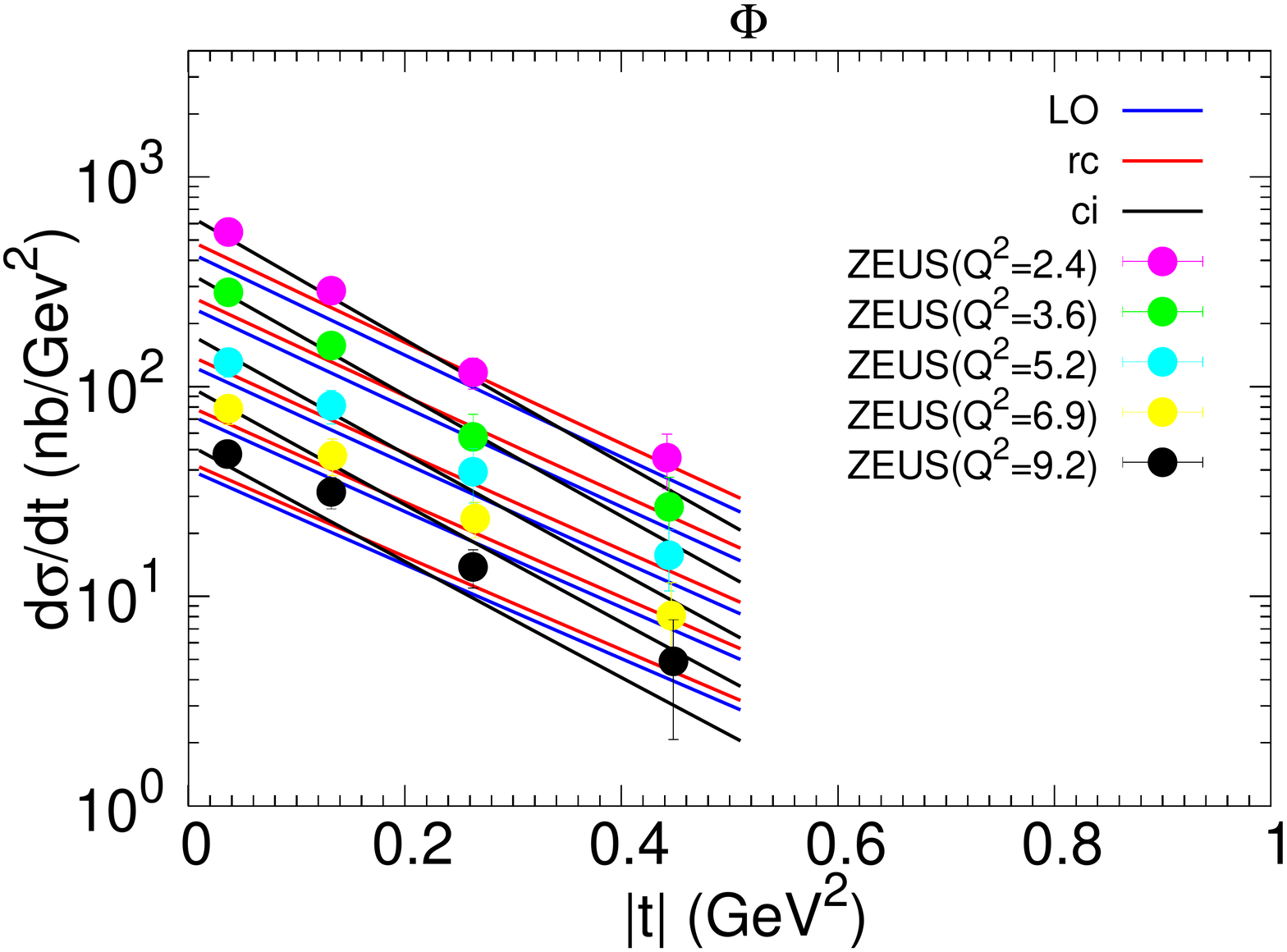}
\end{center}
\vspace{-0.5cm}
\caption{The differential cross section $d\sigma/dt$ for $J/\psi$ and $\phi$ as a function of $t$ at different $Q^{2}$.}
\label{t}
\end{figure}

Figure.~\ref{Q2} shows the elastic cross sections $\sigma$ for $J/\psi$ and $\phi$ productions as a function of photon virtuality $Q^{2}$. The blue, red and black lines represent the results calculating by using the LO, rc and ci dipole amplitudes, respectively (similarly hereinafter). For each of mesons, we consider the experimental data both from H1 and ZEUS Collaborations. For $J/\psi$, one can see that the  higher order dipole amplitudes are good in agreement with the experimental measurement. It can be seen that for $\phi$ production it seems that all the dipole amplitudes give a similarly good description of the data in moderate $Q^{2}$, however only the rc and ci amplitudes can give a precise description of the data in low $Q^{2}$. From Fig.~\ref{Q2} it is almost clear that the NLO amplitudes are more favored by the experimental data.

The elastic cross section $\sigma$ for $J/\psi$ and $\phi$ productions as a function of photon-hadron center of mass energies $W_{\gamma p}$ at different photon virtuality $Q^{2}$ are shown in Fig.~\ref{W}. It is shown from the left panels of Fig.~\ref{W} that the theoretical calculations from the NLO amplitudes are more consistent with the $J/\psi$ data. From the right panels of Fig.~\ref{W}, one can see that the LO calculations have rather poor description of the experimental data, while the NLO computations give a relatively good description of the data although the quality is not as good as the $J/\psi$ case, since the experimental data for $\phi$ meson have large uncertainties. From Fig.~\ref{W}, we can see that the NLO calculations have a better agreement with experimental data than the LO BK equation both for $J/\psi$ and $\phi$.

The differential cross section $d\sigma/dt$ for $J/\psi$ and $\phi$ as a function of the squared momentum transfer $t$ at different photon virtuality $Q^{2}$ are shown in Fig.~\ref{t}. From Fig.~\ref{t}, it seems that the LO and NLO calculations give a similar quality description of the experimental data. It because of the data set is small and with large error bars. However, one can clearly see from the last column in Table~\ref{table:3} that the $\chi^{2}/d.o.f$ computed from the NLO dipole amplitudes are much smaller then the ones from LO cases, which indicate that the NLO corrections take an effective role in the diffractive vector meson productions.

\begin{figure}[b!]
\setlength{\unitlength}{1.5cm}
\begin{center}
\includegraphics[width=0.33\textwidth]{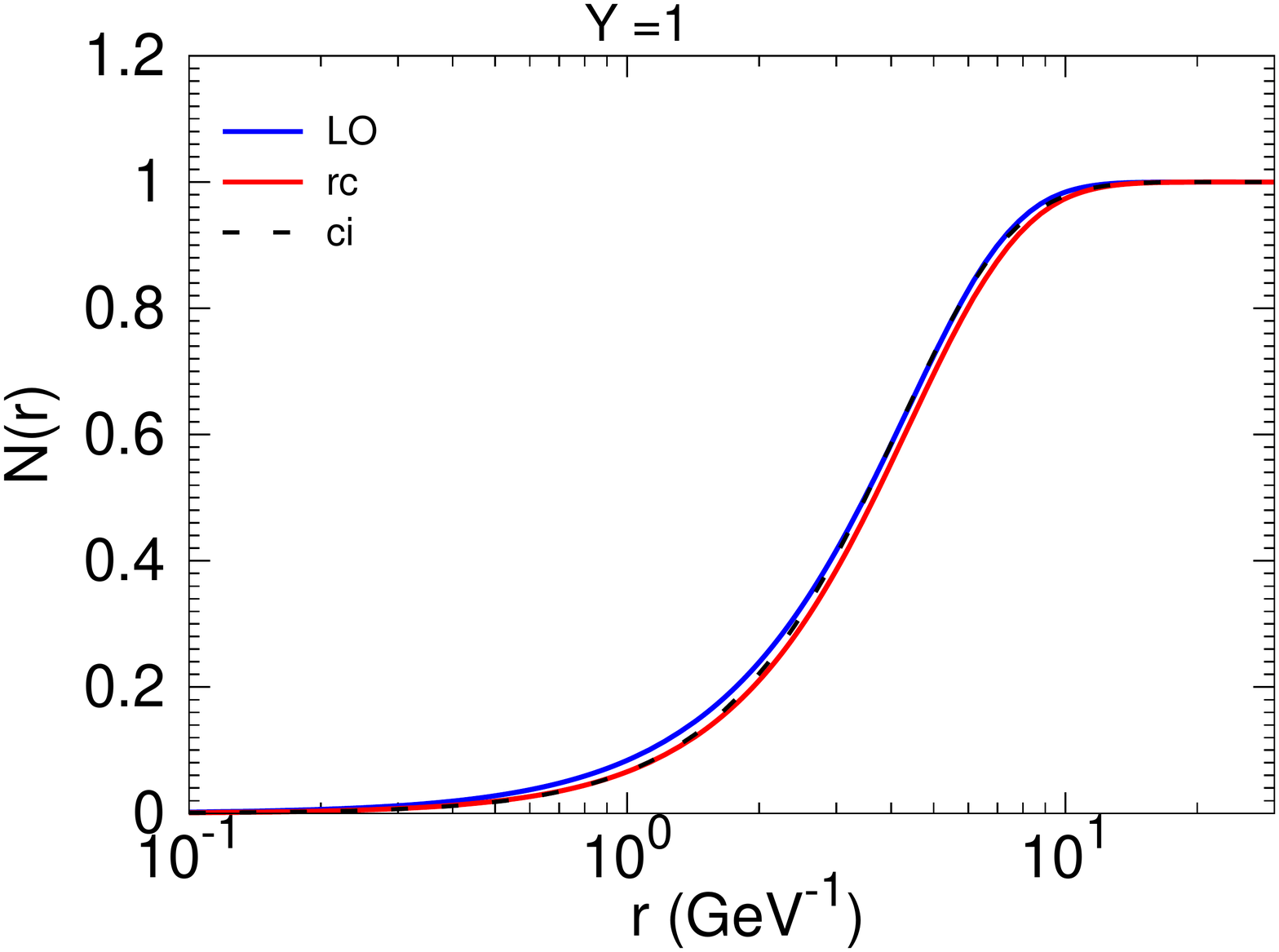}
\vspace{0cm}\hspace{-0.3cm}
\includegraphics[width=0.33\textwidth]{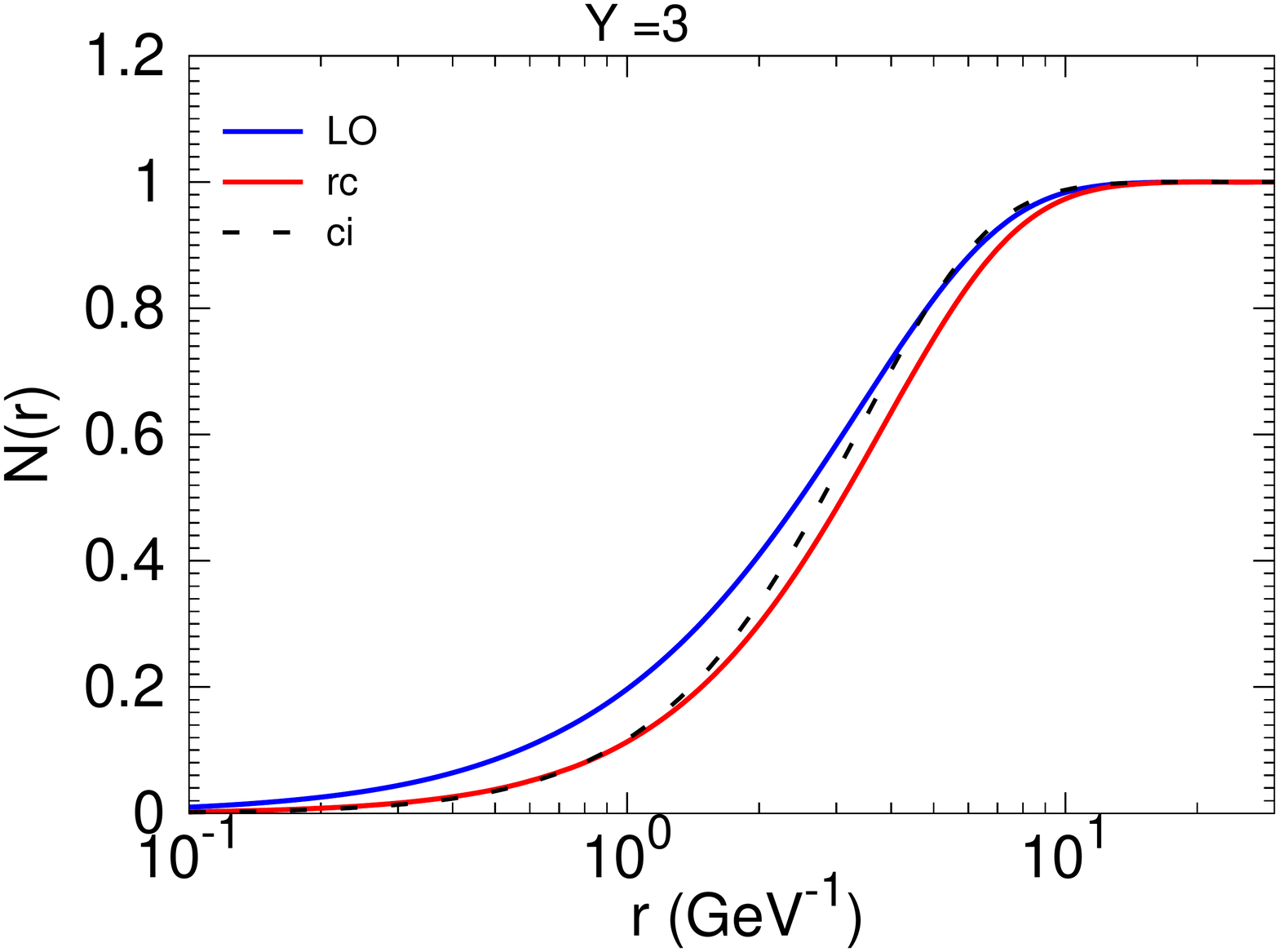}
\vspace{0cm}\hspace{-0.3cm}
\includegraphics[width=0.33\textwidth]{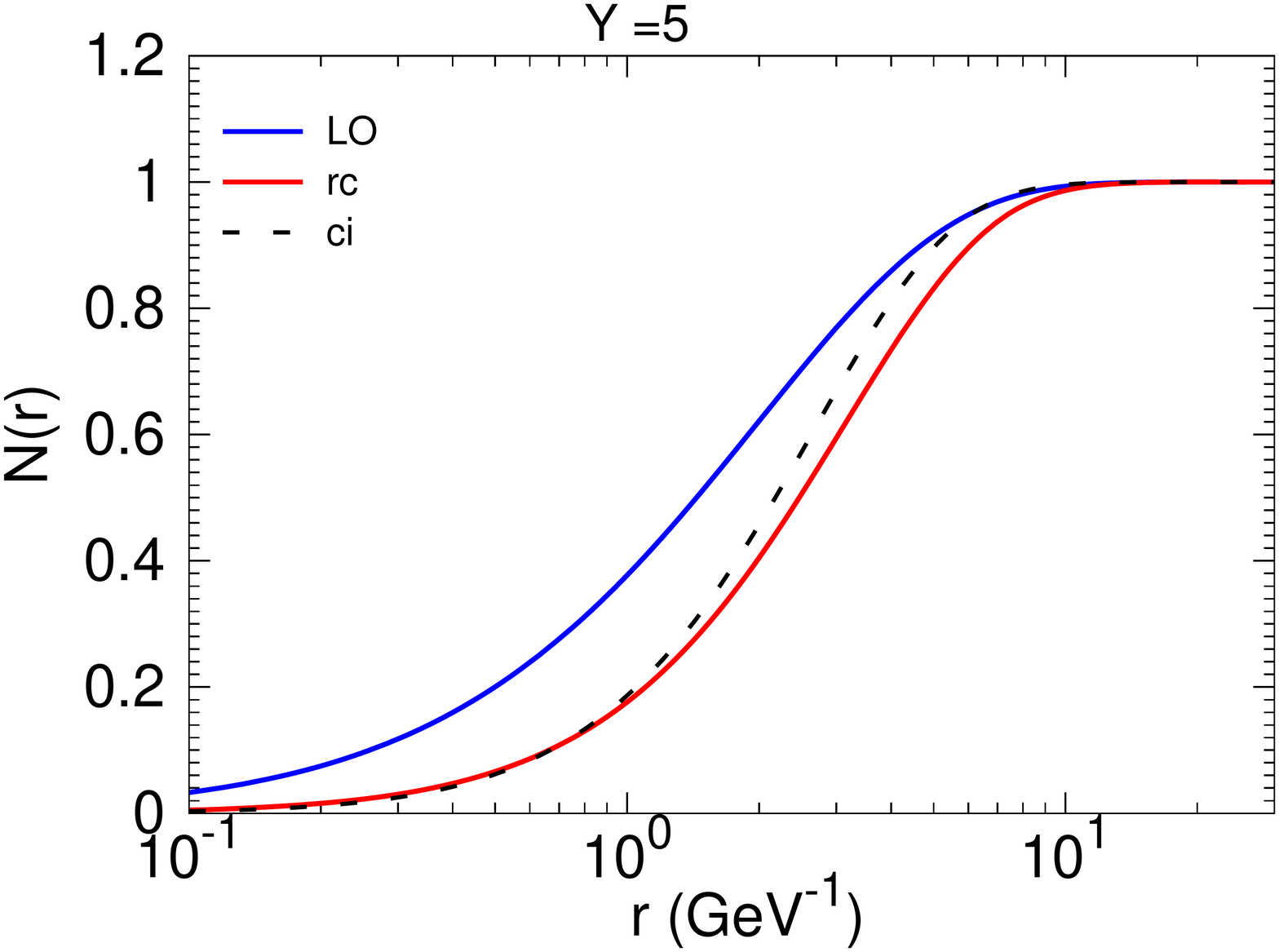}
\end{center}
\vspace{-0.5cm}
\caption{The dipole amplitude for LO, rc and ci BK evolution equations at three different rapidities.}
\label{N_r}
\end{figure}
%

Furthermore, it should be noted that there is not significant difference between the description of the experimental data from the rcBK and ci BK calculations. To better interpret the underlying reasons, we have plotted the dipole amplitude, $\mathcal{N}(r,x)$, as a function of dipole size, $r$, for three different rapidities in Fig.~\ref{N_r}. From Fig.~\ref{N_r}, we find that the difference between the LO and the NLO (rcBK and ci BK) dipole amplitudes is obvious. But, the difference between the amplitudes from rcBK and ci BK equations is tiny up to large rapidities, i.e. $Y=5$. While the largest available rapidity at HERA is about $Y=5$, therefore it is almost impossible to discriminate the NLO running coupling effect from collinear resummations with current HERA data. This is the reason why we cannot see a remarkable difference between $\chi^{2}/d.o.f$ resulting from running coupling and resummation improved dipole amplitudes.

\subsection{Predictions for LHC}
It believes that the experimental data from LHC offer a peculiar way to test the hadronic structure since the higher energy collision will touch even small-$x$ region. As it was shown in Fig.~\ref{N_r}, one can see the difference between the dipole amplitudes from  rc and ci BK equations at the larger rapidities, $Y>5$ (smaller-$x$ region). So, the predictions for the LHC energies are meaningful as higher precision and rapidity data will be released by LHCb collaboration.

In the high energy proton-proton collisions, there are events involving interacting at large impact parameters where the electromagnetic interaction is dominant. In these photon-induced processes, the two protons keep intact after interaction. For the total cross section, it can be written in terms of a convolution of the equivalent photon flux and the photon-proton production cross section. Therefore, the rapidity distribution for the exclusive vector meson production is given by
\be
\label{dif_cross_pp}
\frac{d\sigma[p_1 + p_2 \rightarrow   p_1 \otimes V \otimes p_2]}{dy}
=\Big[\omega \frac{dN_{\gamma/p_1}(\omega)}{d\omega}\sigma_{\gamma p_2 \rightarrow V p_2}(\omega )\Big]_{\omega_L}
+\Big[\omega \frac{dN_{\gamma/p_2}(\omega)}{d\omega}\sigma_{\gamma p_1 \rightarrow V p_1}(\omega )\Big]_{\omega_R},
\ee
where $y$ is the rapidity of the produced vector meson, $\sigma_{\gamma p \rightarrow V p}$ is the total photon-proton cross section, and $\omega$ is the photon energy ($\omega_L= \frac{M_{V}}{2} \exp(-y)$ and $\omega_R= \frac{M_{V}}{2} \exp(y)$). Note that there are two terms in the right hand of the rapidity distribution equation. It is because the photon can be emitted either from left proton or from the right proton.

In Eq.(\ref{dif_cross_pp}), $\frac{dN}{d\omega}$ is the equivalent photon spectrum of the relativistic proton. In the
Weisz$\ddot{a}$cker-Williamsis approximation it can be written as
\be
\label{p flux}
\frac{dN}{d\omega} =  \frac{\alpha_{\mathrm{em}}}{2\pi\omega}\Big[1 + (1 -\frac{2\omega}{\sqrt{s}})^2\Big]
\times\Big( \ln{\xi} - \frac{11}{6} + \frac{3}{\xi}  - \frac{3}{2\xi^2} + \frac{1}{3 \xi^3}\Big),
\ee
where $\xi = 1 + [\,(0.71 \,\mathrm{GeV}^2)/Q_{\mathrm{min}}^2\,]$ with $Q_{\mathrm{min}}^2 \approx (\omega/\gamma_L)^2$ at high energy limit, $\sqrt{s}$ is the proton-proton center of mass energy, $\gamma_L$ is the lorentz factor.

\begin{figure}[t!]
\setlength{\unitlength}{1.5cm}
\begin{center}
\includegraphics[width=0.48\textwidth]{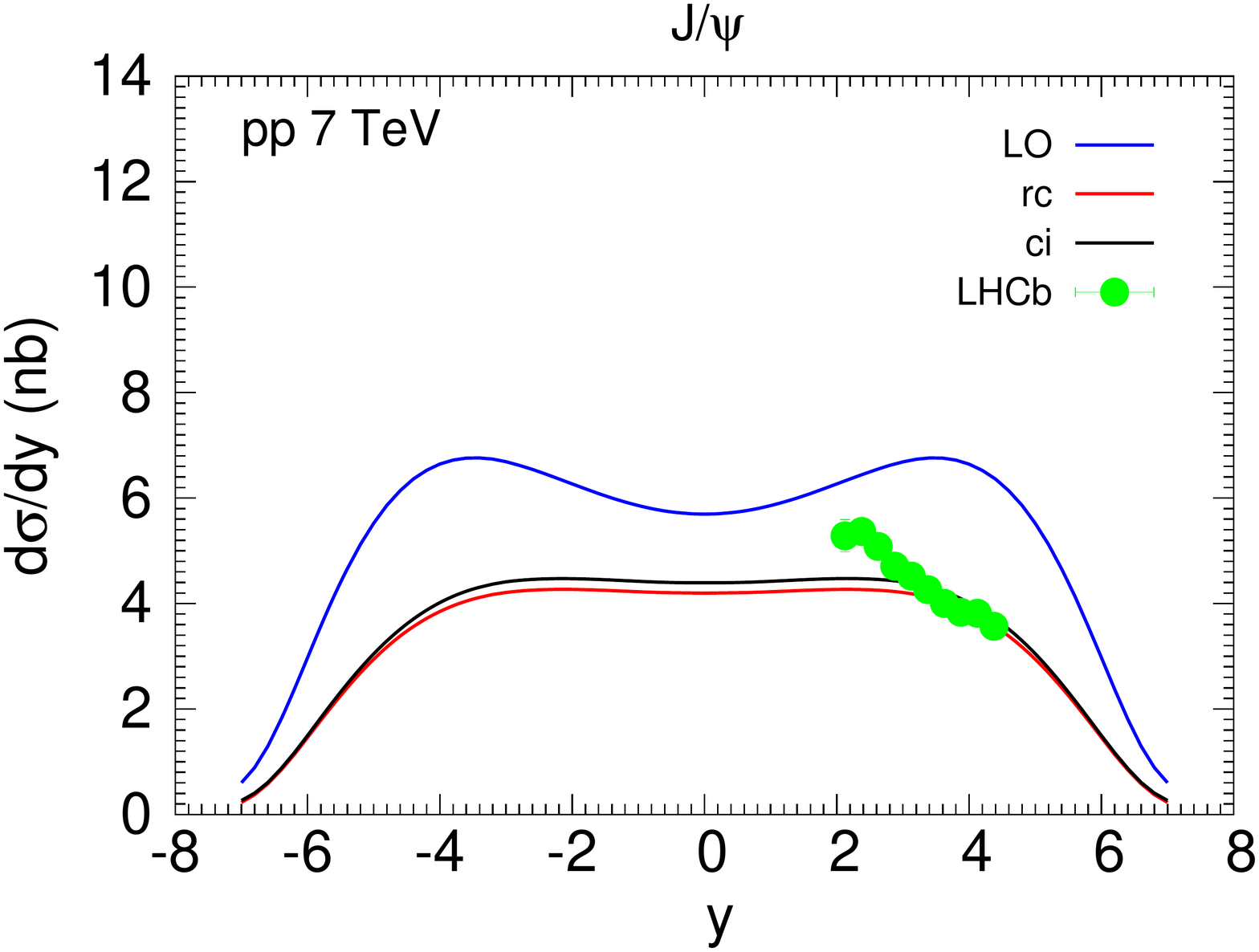}
\vspace{0cm}\hspace{-0.3cm}
\includegraphics[width=0.48\textwidth]{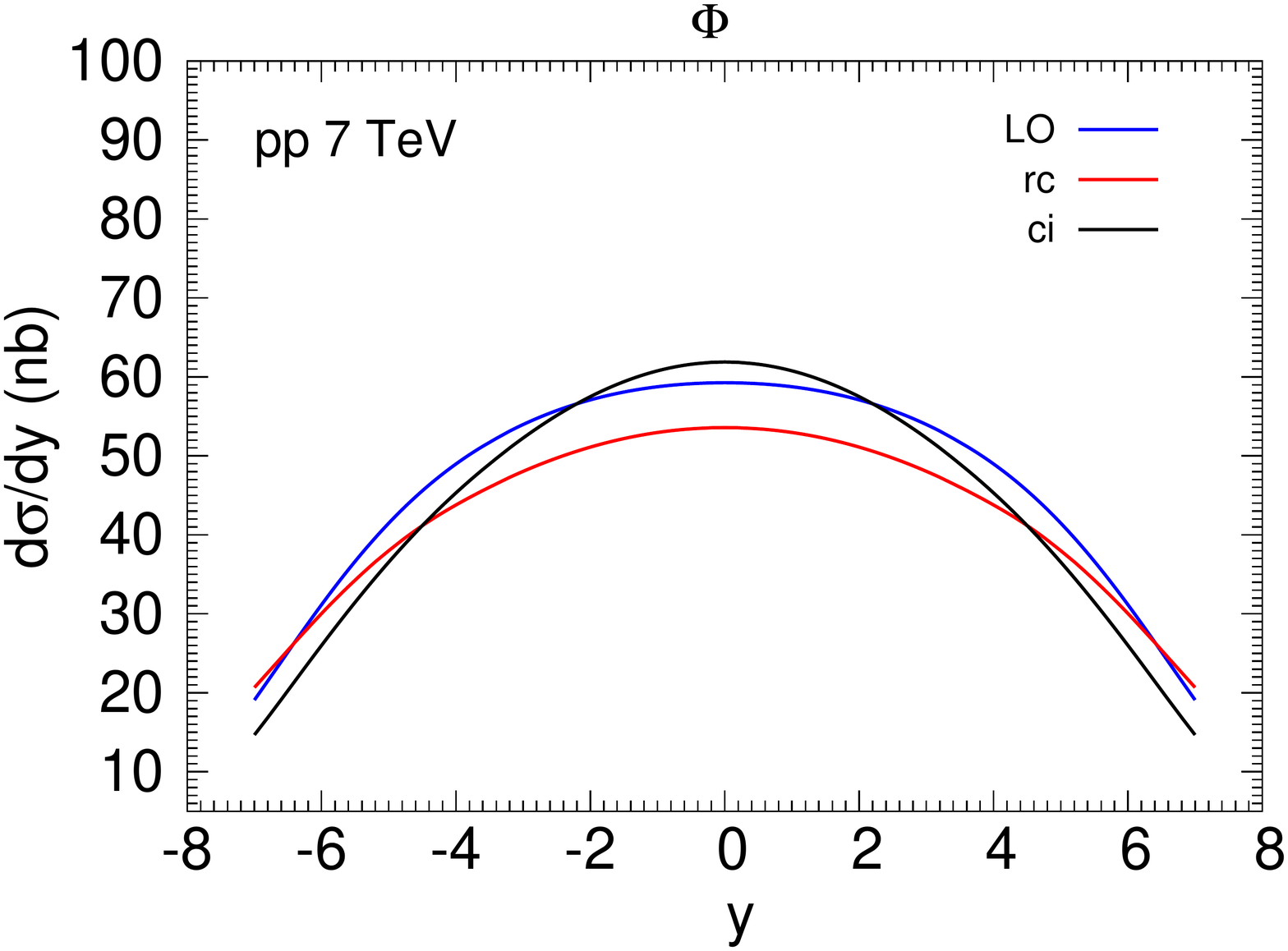}
\end{center}
\vspace{-0.5cm}
\caption{Predictions for the rapidity distributions of $J/\psi$ and $\phi$ mesons in pp collisions at 7 TeV as a function of $y$.}
\label{pp7}
\end{figure}

\begin{figure}[t!]
\setlength{\unitlength}{1.5cm}
\begin{center}
\includegraphics[width=0.48\textwidth]{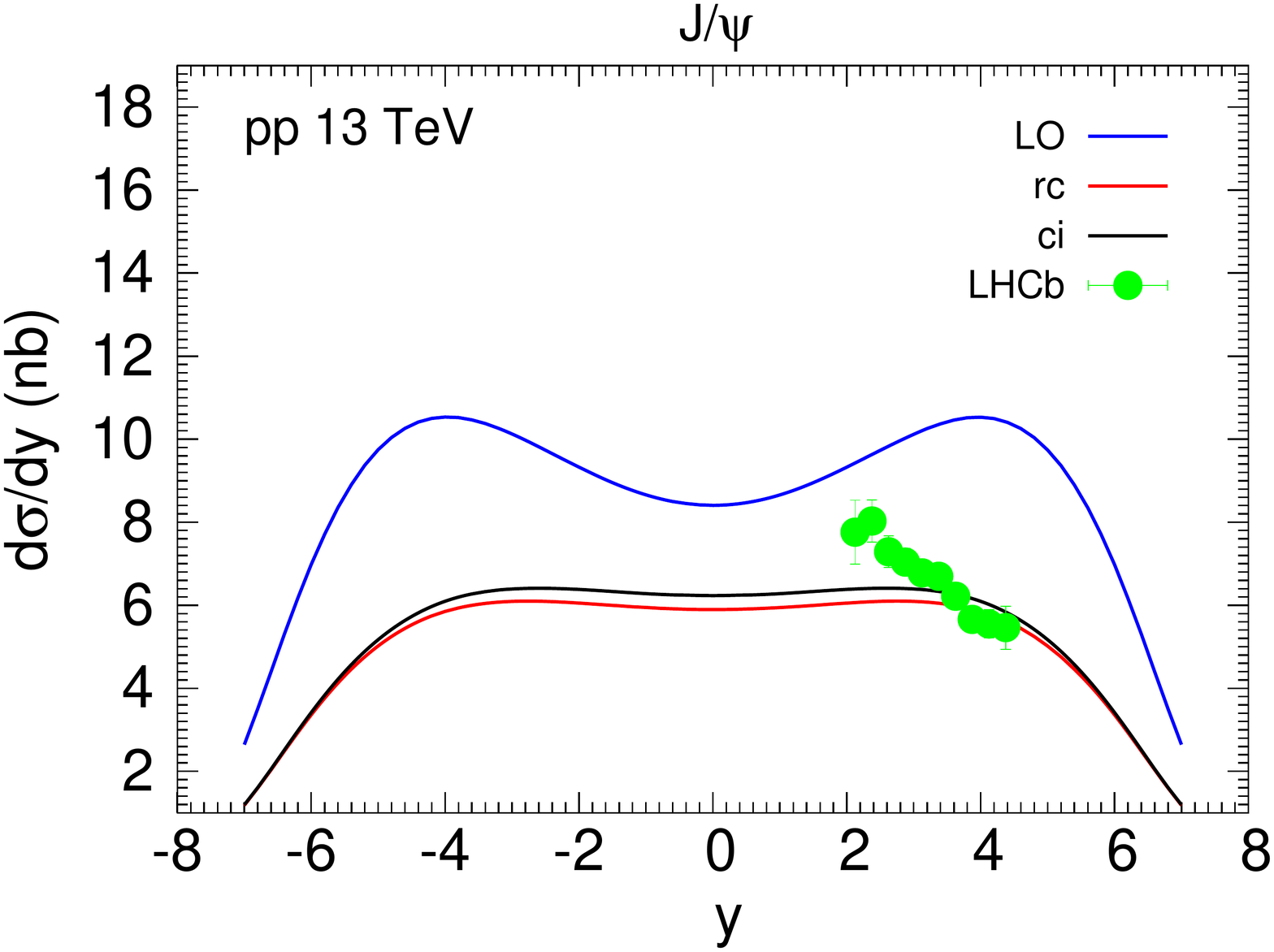}
\vspace{0cm}\hspace{-0.3cm}
\includegraphics[width=0.48\textwidth]{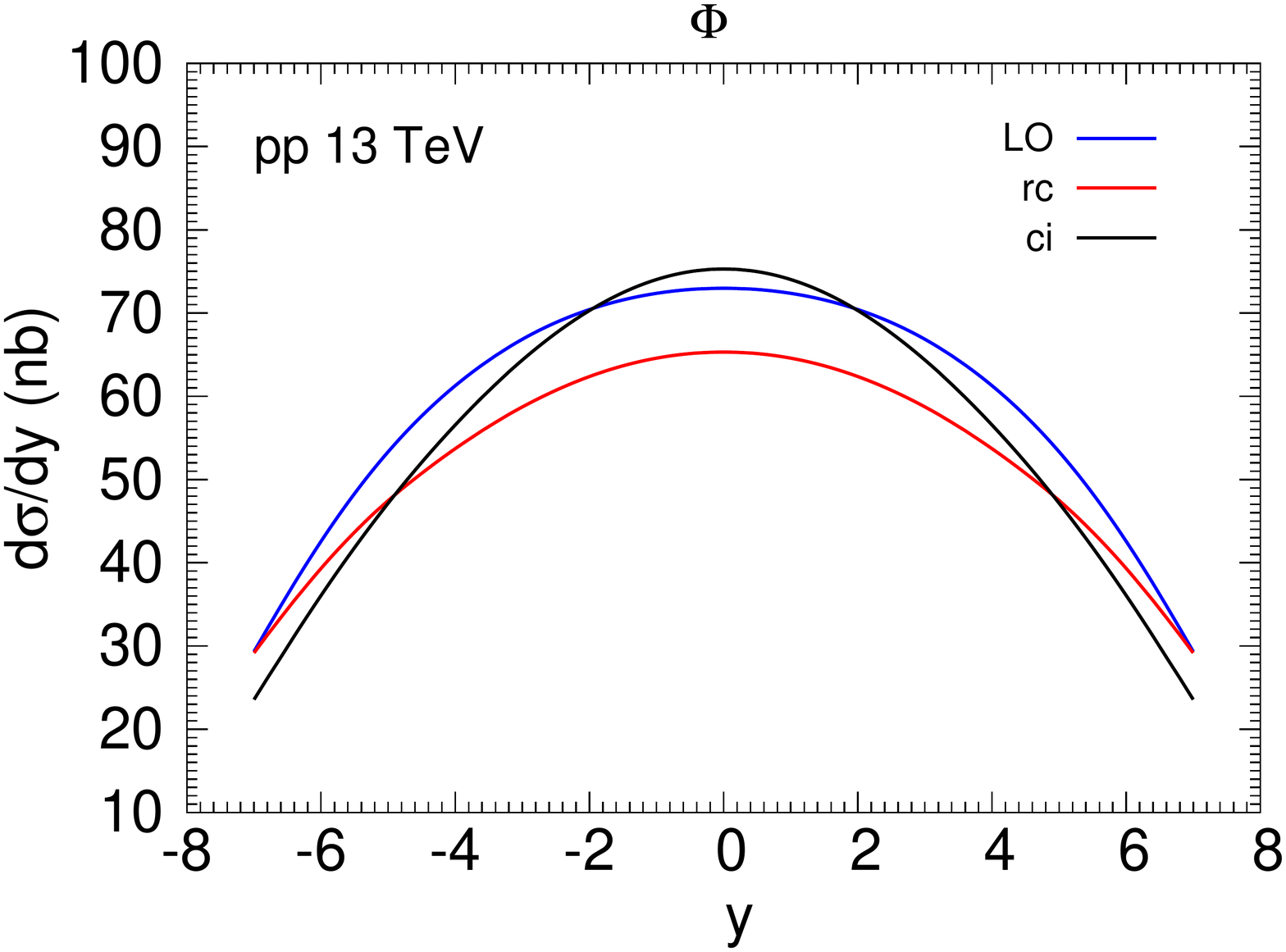}
\end{center}
\vspace{-0.5cm}
\caption{Predictions for the rapidity distributions of $J/\psi$ and $\phi$ mesons in pp collisions at 13 TeV as a function of $y$ .}
\label{pp13}
\end{figure}

Moreover, the total photon-proton cross section $\sigma_{\gamma p \rightarrow V p}$ can be integrated from the differential cross section in Eq.(\ref{dif_cross}). The integral over $t$ can be rewritten as follows
\be
\label{dif_to_tot}
\sigma_{\gamma p \rightarrow V p} =  \int_{-\infty}^{0}\frac{d\sigma^{\gamma p\rightarrow Vp}}{dt}dt.
\ee
Using the above formalism and the parameters obtained from fitting the HERA data, we can predict the rapidity distributions for diffractive $J/\psi$ and $\phi$ productions in proton-proton collisions at LHC energies. Figures.~\ref{pp7} and \ref{pp13} show our predictions for the rapidity distributions of exclusive $J/\psi$ and $\phi$ in proton-proton collisions at 7 TeV and 13 TeV, respectively. For the LHC kinematics region, there are possible rapidities whose corresponding Bjorken-$x$ can be larger then $x_0$ for one of the proton but still smaller then $x_0$ for the other proton. In order to get a smooth curve, we make a linear extrapolation for the dipole amplitudes when $ x > 0.01$. We consider three kinds of dipole amplitudes (LO, rc, ci amplitudes) to calculate the rapidity distributions for exclusive vector meson productions and compare with the released data from LHCb\cite{Aaij,Aaij2}. The numerical results in Figs.~\ref{pp7} and \ref{pp13} show that the NLO dipole amplitudes give a better agrement with experimental data points as expected. For completeness, we present our predictions of the total cross section with different kinds of dipole amplitudes in Tables~\ref{table:4} and \ref{table:5}. From the tables, one can see that the production rates of the vector mesons ($J/\psi$ and $\phi$) are suppressed by the NLO effect, which satisfy with the theoretical execrations.

\begin{table}[h!]
  \begin{center}
  \begin{tabular}{cc|ccc}
  \hline
  &  &\quad  LO    &\quad   rc   &\quad  ci  \\
  \hline
 & 7 TeV &\quad  37.181 nb  &\quad   23.204 nb  &\quad 24.229 nb \\
  \hline
 & 13 TeV &\quad  61.217 nb  &\quad   31.016 nb &\quad  37.608 nb  \\
  \hline
  \end{tabular}%
  \end{center}
  \caption{The $J/\psi$ total cross section with different dipole evolution equations in pp collisions.}
  \label{table:4}
\end{table}%
\begin{table}[h!]
  \begin{center}
  \begin{tabular}{cc|ccc}
  \hline
  &  &\quad  LO    &\quad   rc   &\quad  ci  \\
  \hline
 & 7 TeV &\quad  331.567 nb   &\quad   301.806 nb  &\quad  317.551 nb \\
  \hline
 & 13 TeV &\quad  419.120 nb  &\quad   374.290 nb &\quad  398.908 nb \\
  \hline
  \end{tabular}%
  \end{center}
  \caption{The $\phi$ total cross section with different dipole evolution equations in pp collisions.}
  \label{table:5}
\end{table}%
%
In conclusion, we have investigated the exclusive vector meson photoproduction for $J/\psi$ and $\phi$ at HERA in the framework of color glass condensate. By comparing the results from the rcBK and ci BK equations with those from the LO BK equation, we find that the results from NLO equations are more consistent with experimental data than LO BK equation. We also present our predictions for the rapidity distributions in pp collisions by using parameters obtained from fitting the HERA data. These results indicate that the NLO effects are significant in the calculation of vector meson production at LHC energies. Furthermore, the higher order corrections considered in this work are part of NLO corrections to the BK evolution equation. As we have studies in Ref.\cite{Xiang2} that the rare fluctuations also have a large corrections to the evolution equation once the gluon loop contributions are included into the rcBK equation. Therefore, the exclusive vector meson production with a rare fluctuation corrections are worth to explore in the next work.

\vspace*{.5cm}


\begin{acknowledgments}
 This work is supported by the National Natural Science Foundation of China under Grant Nos.11947119, 11765005, 11305040, 11847152, and 11775097; the Fund of Science and Technology Department of Guizhou Province under Grant Nos.[2018]1023, and [2019]5653; the Education Department of Guizhou Province under Grant No.KY[2017]004; Qian Kehe Platform Talents No. [2017]5736-027; the National key research and development program of China under Grant 2018YFE0104700, and the Grant CCNU18ZDPY04; and the 2018 scientific research startup foundation for the introduced talent of Guizhou University of Finance and Economics under grant No. 2018YJ60.

\end{acknowledgments}



\end{document}